\RequirePackage[bookmarksnumbered,unicode]{hyperref}
\documentclass[sigconf,screen]{acmart}

\usepackage{url}
\usepackage[most]{tcolorbox}
\usepackage{hyperref}
\usepackage{multirow}
\usepackage{subcaption}
\usepackage{enumitem}
\usepackage{algorithmic}
\usepackage{algorithm}
\usepackage{booktabs}
\usepackage{graphicx} 
\usepackage{xspace}
\usepackage{caption}
\usepackage{amsmath,amsfonts}
\usepackage{textcomp}
\usepackage{xcolor}
\usepackage{pifont}
\usepackage{bm}

\setlist[enumerate]{nolistsep}

\newcommand{\tool}{SEER\xspace}

\newcommand{\gsz}[1]{\textcolor{black}{{#1}}}

\usepackage{tikz}
\newcommand*{\circled}[1]{\lower.7ex\hbox{\tikz\draw (0pt, 0pt)%
    circle (.5em) node {\makebox[1em][c]{\small #1}};}}
    
\AtBeginDocument{%
  \providecommand\BibTeX{{%
    \normalfont B\kern-0.5em{\scshape i\kern-0.25em b}\kern-0.8em\TeX}}}

\setcopyright{none}
\renewcommand\footnotetextcopyrightpermission[1]{}

\begin{document}
\title{SEER: Enhancing Chain-of-Thought Code Generation through Self-Exploring Deep Reasoning}



\author{Shuzheng Gao}
\affiliation{%
  \institution{The Chinese University of Hong Kong}
  \city{Hong Kong}
  \country{China}}
\email{szgao23@cse.cuhk.edu.hk}

\author{Chaozheng Wang}
\affiliation{%
  \institution{The Chinese University of Hong Kong}
  \city{Hong Kong}
  \country{China}}
\email{czwang23@cse.cuhk.edu.hk}

\author{Cuiyun Gao}
\authornote{Corresponding author.}
\affiliation{%
  \institution{Harbin Institute of Technology}
  \city{Shenzhen}
  \country{China}}
\email{gaocuiyun@hit.edu.cn}

\author{Michael R. Lyu}
\affiliation{%
  \institution{The Chinese University of Hong Kong}
  \city{Hong Kong}
  \country{China}}
\email{lyu@cse.cuhk.edu.hk}

\begin{abstract}
Code generation, the task of creating executable programs from natural language requirements, has recently seen tremendous advances through
Chain-of-Thought (CoT) reasoning, which enables {Large Language Models (LLMs)}
to develop high-level {reasoning}
plans before writing code. Recent research has proposed various methods to enhance models' CoT reasoning for code generation such as prompt engineering and supervised fine-tuning. However, existing approaches still face three critical limitations: (1) limited exploration of diverse reasoning paths, which constrains generalization across various {programming}
scenarios, (2) lack of quality assessment for intermediate reasoning steps, which hampers the reliability of the generated plans and code, and (3) the potential negative impact of ``overthinking'', potentially leading to unnecessarily complex and
incorrect solutions. 
To address these limitations, we frame CoT code generation as a decision making problem and present \tool, a \textbf{SE}lf-\textbf{E}xploring deep \textbf{R}easoning framework that enables accurate and adaptive reasoning for code generation.
\tool introduces three key components: (1) Diverse reasoning path exploration, which aims at exploring diverse reasoning paths and annotating intermediate steps without relying on manual experts or closed-source  proprietary models; (2) Reasoning quality-aware model training, which trains a policy model for generating candidate reasoning steps and a value model for assessing their quality; 
and (3) Adaptive CoT reasoning, which dynamically switches
between direct generation and step-by-step reasoning for different problems. Experiments on state-of-the-art code LLMs DeepSeek-Coder and Qwen2.5-Coder demonstrate that \tool achieves remarkable performance gains across three popular code generation benchmarks, consistently outperforming all baseline methods and achieving absolute improvements by 4.2\% $\sim$ 9.3\% in MBPP, 1.9\% $\sim$ 9.1\% in HumanEval and 3.5\% $\sim$ 5.3\% in LiveCodeBench, respectively. 
\end{abstract}

\maketitle

\footnote{The paper was completed in Feb. 2025, submitted to ICSE 2026 in Mar. 2025, received a major revision in Jun. 2025, and was finally accepted in Oct. 2025.}

\section{Introduction}\label{sec:intro}

Code generation aims at automatically generating source code that adheres to a given programming problem~\cite{DBLP:conf/icse/JainVINPR022,DBLP:conf/icse/LiLLJHH23}. 
As software systems become increasingly complex and the demand for skilled developers continues to increase, automated code generation {holds great potential} 
to advance the software development process by reducing repetitive manual programming efforts and improving productivity~\cite{DBLP:journals/corr/abs-2303-08774,DBLP:journals/corr/abs-2407-21783,DBLP:conf/icse/GaoMG000L24}. Recent advances in Large Language Models (LLMs) have driven considerable progress in software engineering tasks~\cite{DBLP:journals/tosem/GaoGHZNXL23,DBLP:journals/corr/abs-2501-01329,li2023structured,DBLP:conf/icml/0003W0D024, DBLP:conf/acl/WenYGXY25, DBLP:journals/corr/abs-2404-15596}, such as ChatGPT~\cite{ChatGPT} and Deepseek-Coder~\cite{guo2024deepseek}. 
Despite these success cases, producing accurate code for complex problems remains an important challenge that requires sophisticated reasoning to align implementation with problem specifications~
\cite{DBLP:journals/tosem/JiangDWFSLJJ24,DBLP:conf/kbse/ZhangCW024}.

\begin{figure}[t]
    \centering
    \includegraphics[width=0.9\linewidth]{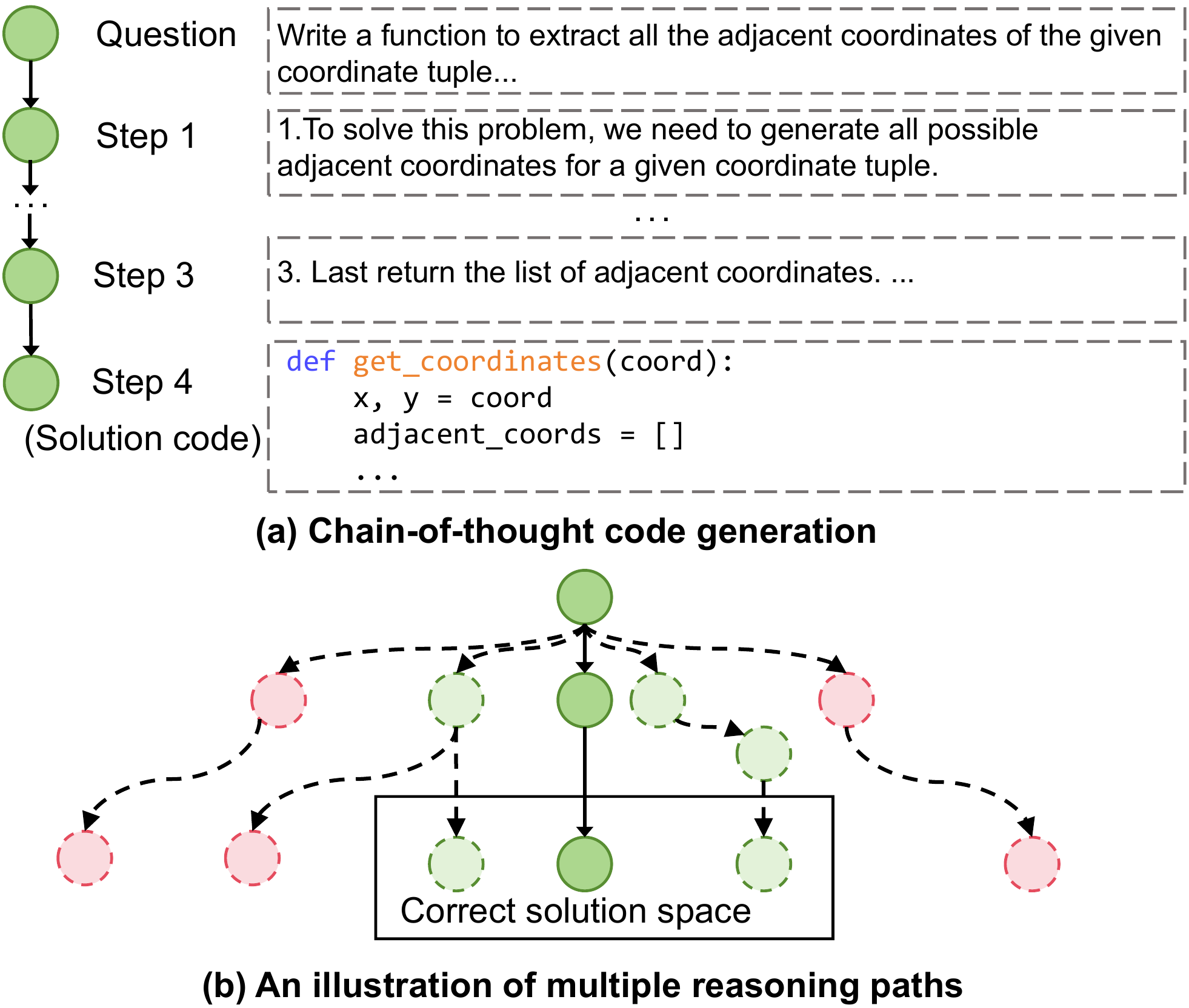}
    \caption{An illustration of chain-of-thought code generation (a) and multiple reasoning paths (b). The green and red node denote correct and incorrect steps, respectively. The dashed lines and circles represent unexplored paths.}
    \label{fig:illustration}
    \vspace{-0.45cm}
\end{figure}

Recently, Chain-of-Thought (CoT)~\cite{DBLP:conf/nips/Wei0SBIXCLZ22,DBLP:conf/nips/KojimaGRMI22} has demonstrated great potential in improving the reasoning abilities of LLMs for complex tasks by breaking problems into intermediate steps and solving them step by step. As shown in Figure~\ref{fig:illustration} (a),
CoT enhances code generation by enabling models to explicitly outline a high-level reasoning
plan before producing the code, which has been shown to be effective~\cite{DBLP:journals/tosem/JiangDWFSLJJ24}. Recent research has explored various {CoT} methods {for}
code generation, with some leveraging prompt engineering and in-context learning to elicit reasoning chains during inference time~\cite{DBLP:journals/tosem/JiangDWFSLJJ24,li2023structured,DBLP:conf/icse/Gao0GL25,DBLP:conf/kbse/GaoWGWZL23}. {However, these techniques rely on large proprietary models and struggle with smaller models, as the latter typically lack the capacity to learn reasoning processes from prompts alone.}
To address this limitation, recent studies have explored distilling high-quality reasoning chains from advanced proprietary models such as GPT-4~\cite{DBLP:journals/corr/abs-2303-08774} and using these curated reasoning data to train lightweight language models capable of performing CoT reasoning~\cite{DBLP:journals/tse/YangZCZZC24}.

\begin{figure}[t]
    \centering
    \includegraphics[width=1\linewidth]{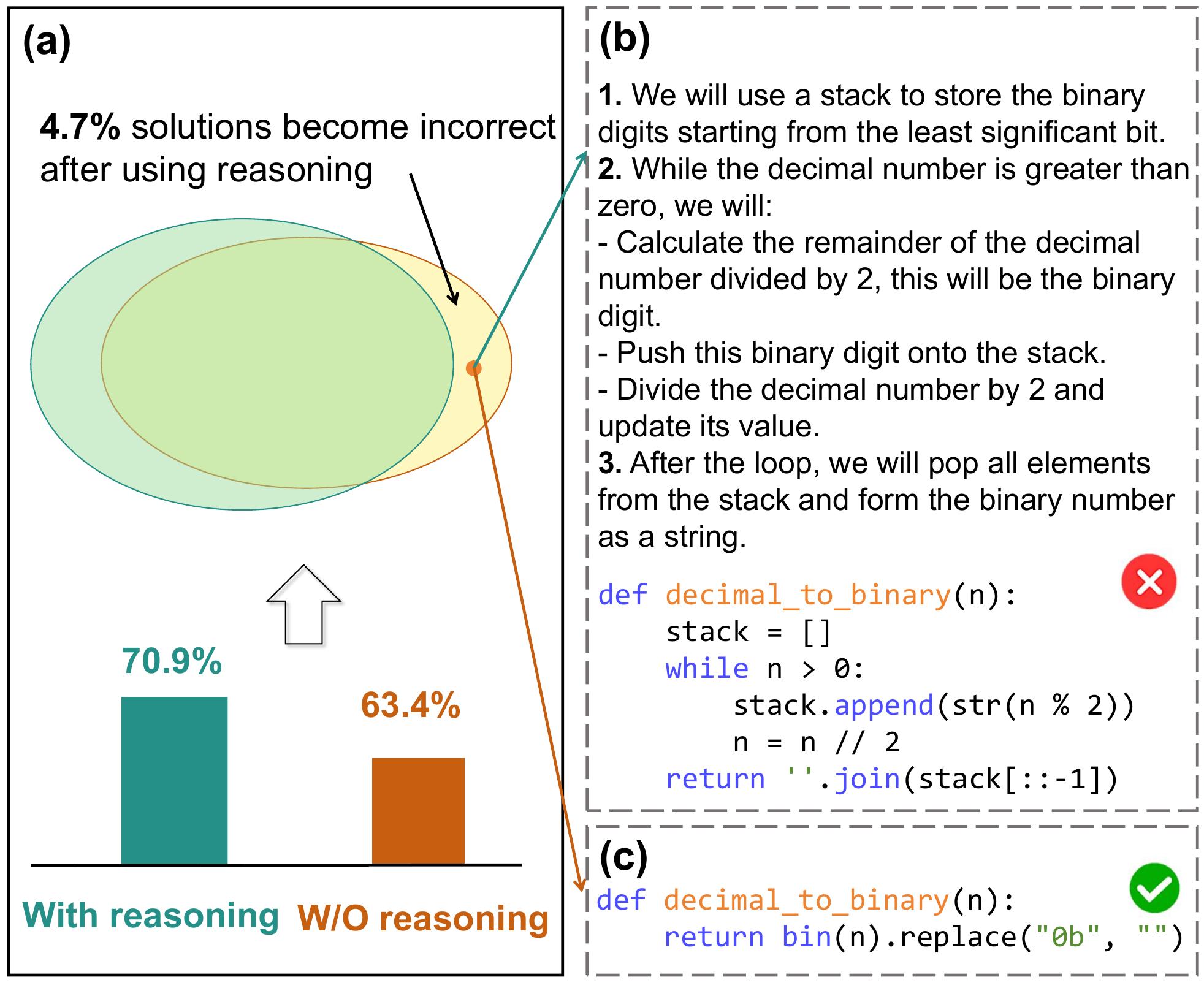}
    \caption{An example of ``overthinking'' in code generation: {Reasoning improves overall accuracy but can also lead to errors (a).} While solution (b) includes detailed reasoning, it opts for a complex implementation and fails to handle zero input. In contrast, solution (c) takes a direct approach using the built-in \texttt{bin()} function and correctly handles all inputs.}
    \label{fig:motivation}
    \vspace{-0.5cm}
\end{figure}

Despite these advances, current CoT code generation approaches still face {the following}
critical limitations: (1) First, {they}
\textbf{lack effective exploration of diverse reasoning paths}. Although programming problems often permit multiple valid {solutions,} 
as shown in Figure~\ref{fig:illustration} (b), existing {approaches}
typically rely on training data that contain only a single reasoning path {(indicated by the solid line)} due to the high cost of distilling proprietary models. This leads to models overfitting to specific reasoning patterns, constraining their ability to generalize across various problem-solving scenarios.
(2) Second, these {approaches}
\textbf{lack quality assessment of intermediate reasoning steps}. Generating reasoning based on flawed intermediate steps often leads
to incorrect final solutions. Existing approaches solely train models with correct reasoning paths, ignoring the importance of step-level correctness. The oversight prevents models from learning to evaluate reasoning quality and prioritize promising paths.
{However, collecting step-level correctness data presents significant challenges. While final solution quality can be validated through test cases, automatically evaluating intermediate reasoning steps remains difficult, as even incorrect solutions may contain valid intermediate reasoning components.}
(3) Third, previous work has \textbf{overlooked the negative impact of ``overthinking''}. Although CoT reasoning generally improves overall performance,we find that it can also lead models to pursue unnecessarily complex and
incorrect {solutions}
that they are not familiar with, even for relatively simple problems. For example, as shown in Figure~\ref{fig:motivation} (a), although CoT reasoning improved DeepSeek-Coder-6.7B-Instruct's performance on MBPP~\cite{austin2021program} by 7.5\%, it also caused 4.7\% of previously correct answers to become incorrect (details referring to Section \ref{sec:ablation}). This not only results in performance degradation on problems that could have been straightforwardly solved  (e.g. Figure~\ref{fig:motivation} (b) and (c)) but also increases computational costs through higher token consumption.

To address these limitations, we frame CoT code generation as a decision-making problem and present \tool, a \textbf{SE}lf-\textbf{E}xploring deep \textbf{R}easoning framework that enables accurate and adaptive reasoning for code generation.
\tool  contains three stages: 
1) Diverse reasoning path exploration, which aims at exploring diverse reasoning paths and annotating intermediate steps by customizing
Monte Carlo Tree Search 
with reasoning path perturbation and refinement, eliminating the need for manual experts or proficiency models;
2) Reasoning quality-aware model training,
which trains a policy model for generating candidate reasoning steps and a value model for evaluating their quality using self-explored data; 
{3) Adaptive CoT reasoning, which dynamically selects between direct generation and step-by-step reasoning for different problems.} 

To evaluate the effectiveness of \tool, we conduct extensive experiments on two state-of-the-art code LLMs, including DeepSeek-Coder~\cite{guo2024deepseek} and Qwen2.5-Coder~\cite{hui2024qwen2}. Experimental results demonstrate that \tool outperforms all baseline methods such as DeepSeek-R1-Zero's reinforcement learning~\cite{guo2025deepseek} and improves the base model by 4.2\% $\sim$ 9.3\%, 1.9\% $\sim$ 9.1\%, and 3.5\% $\sim$ 5.3\% in terms of pass@1 on three popular coding benchmarks: MBPP~\cite{austin2021program}, HumanEval~\cite{chen2021evaluating}, and LiveCodeBench~\cite{jain2024livecodebench}, respectively. Furthermore, when enhanced with \tool, DeepSeek-Coder-6.7B outperforms models with 4-10 times more parameters, such as Llama3-70B~\cite{DBLP:journals/corr/abs-2407-21783} and DeepSeek-Coder-33B~~\cite{guo2024deepseek}. Furthermore, Qwen2.5-Coder-7B+\tool achieves competitive results compared to some closed-source proprietary models, including GPT-4-Turbo~\cite{DBLP:journals/corr/abs-2303-08774} and Gemini 1.5 Flash 002~\cite{DBLP:journals/corr/abs-2312-11805}.

In general, we make the following contributions:
\begin{enumerate}
\item {To the best of our knowledge, we are the first to frame CoT code generation as a decision-making problem and enable autonomous exploration of step-annotated reasoning paths without relying on manual expert annotations or proprietary models.} 

\item {We propose \tool, a framework that enhances code generation through diverse reasoning path exploration, reasoning quality-aware model training, and adaptive CoT reasoning.}

\item Extensive experimentation demonstrate that \tool improves the performance of state-of-the-art LLMs by a large margin across multiple popular benchmarks and enables smaller LLMs to outperform larger ones.


\item To facilitate further research, we publicly release the code, trained model, and step-annotated dataset on GitHub~\cite{replicate}.
\end{enumerate}

\section{Preliminary}\label{sec:back}

\subsection{Problem Formulation}
We formulate chain-of-thought code generation as a decision-making problem using the Markov Decision Process (MDP)~\cite{feinberg2012handbook}. In the MDP framework, an agent begins from an initial state $\mathbf{s}_0$ and takes sequential actions $\mathbf{a}_t$, transitioning from state $\mathbf{s}_{t-1}$ to $\mathbf{s}_t$ at each timestep until reaching a final state $\mathbf{s}_n$ that contains the solution. For code generation, states and actions are represented as token sequences, with the initial state $\mathbf{s}_0={x_0,...,x_L}$ being the input prompt, where $L$ denotes the prompt length. Following~\cite{DBLP:journals/tosem/JiangDWFSLJJ24} and as illustrated in Figure~\ref{fig:illustration}, we use a single implementable sub-task to solve this question as a step and the last step will give the final solution code. 
At each timestep $t$, given the prompt and current steps as state $\mathbf{s}_t=(\mathbf{s}_0,\mathbf{a}_1,...,\mathbf{a}_{t-1})$, a new step is generated as action $\mathbf{a}_t=(y_{t,0},...,y_{t,n_t})$, where $y_{t,i}$ is a single token and {$n_t$ is the length of the action's content}. The state transition function deterministically updates the state through token concatenation: $\mathbf{s}_{t+1}=\text{Cat}(\mathbf{s}_t, \mathbf{a}_t)$.

Given the large action space in the LLM's step-by-step generation process~\cite{guo2025deepseek},
our primary goal is to develop two models: a step-level policy model $\pi_\theta(s)$ and a value model $V_\phi(s)$. The policy model generates some promising actions at each state, while the value model assesses the expected returns from the current partial solution to guide the LLM in selecting the most reasonable subsequent reasoning step. Therefore, to train these models, we need to collect diverse data that contain both correct and incorrect solution paths, with annotations for intermediate reasoning steps that include both action $\mathbf{a}$ and value $\mathbf{v}$ indicating action quality.

\begin{figure}[t]
    \centering
    \includegraphics[width=0.7\linewidth]{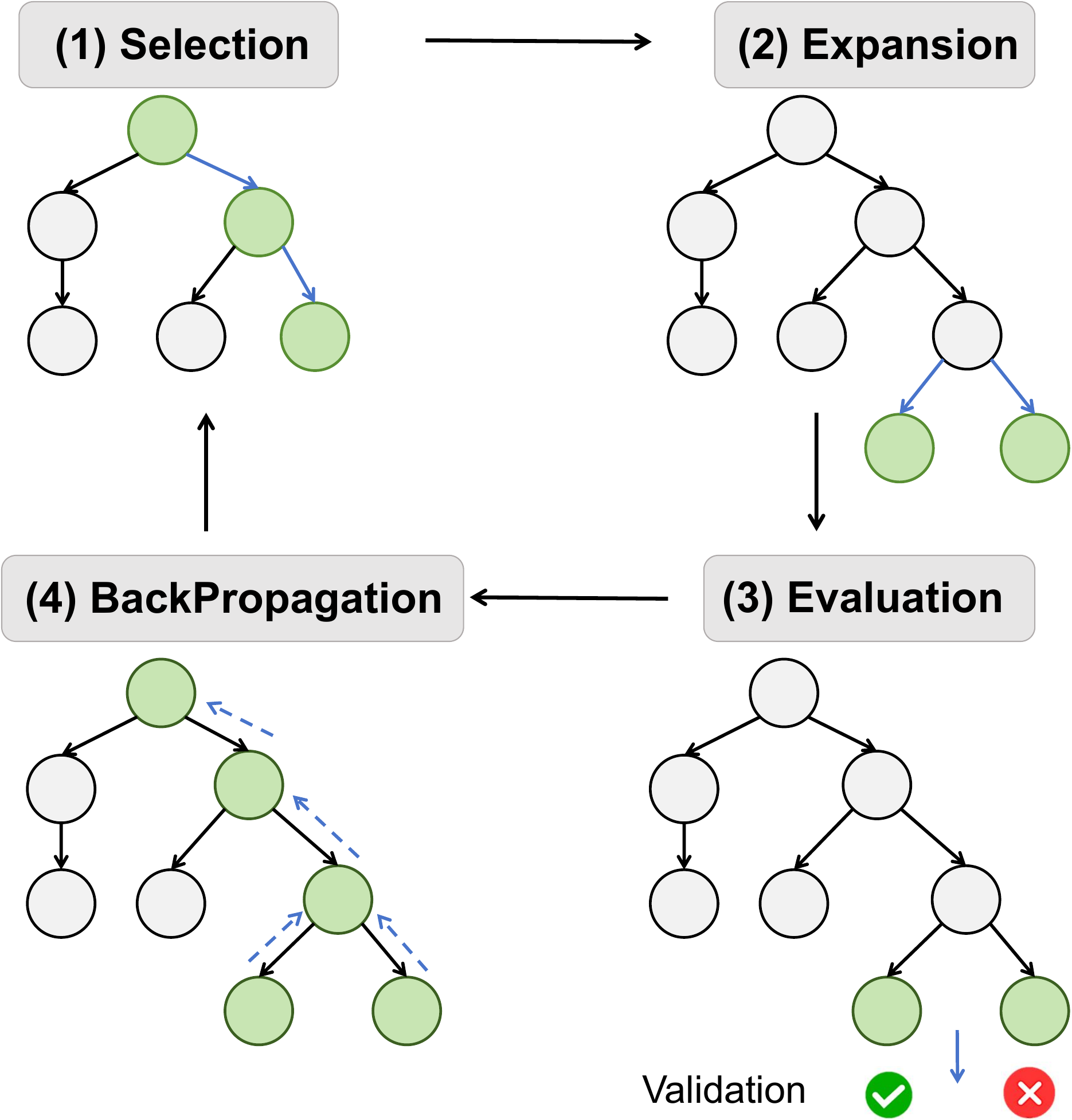}
    \caption{An overview of the four key operations in MCTS. 
    }
    \label{fig:mcts}
    \vspace{-0.3cm}
\end{figure}
\subsection{Monte Carlo Tree Search}
Monte Carlo Tree Search (MCTS) is a heuristic search algorithm that has gained noticeable attention since its breakthrough application in AlphaGo~\citep{silver2016mastering}. It combines tree search with Monte Carlo sampling, offering a powerful approach for sequential decision-making problems, especially those with large state spaces where traditional methods such as the minimax search become computationally intractable~\citep{silver2017mastering,coulom2006efficient}. As shown in Figure~\ref{fig:mcts}, MCTS iteratively builds a search tree through four key steps: (1) \textbf{Selection}: traversing the tree from root to leaf and selecting promising nodes to expand; 
(2) \textbf{Expansion}: adding new child nodes to expand the tree based on available actions in the current state; (3) \textbf{Evaluation}: estimating the value of leaf nodes through random rollouts to terminal states or neural network value approximation; and (4) \textbf{Backpropagation}: updating node statistics (visit counts and action values) along the traversed path to improve future selection. Recent research has extended MCTS beyond its traditional gaming domain, exploring its potential for LLM reasoning~\cite{bai2024digirl,best2019dec}. Specifically, researchers have demonstrated success in using MCTS to simulate expert-like problem solving and annotate step-by-step reasoning processes for tasks such as mathematical reasoning and proof generation~\citep{DBLP:journals/corr/abs-2405-03553,DBLP:journals/corr/abs-2405-00451}.


\section{Proposed Approach}\label{sec:approach}
\begin{figure*}[t]
    \centering
    \includegraphics[width=0.8\linewidth]{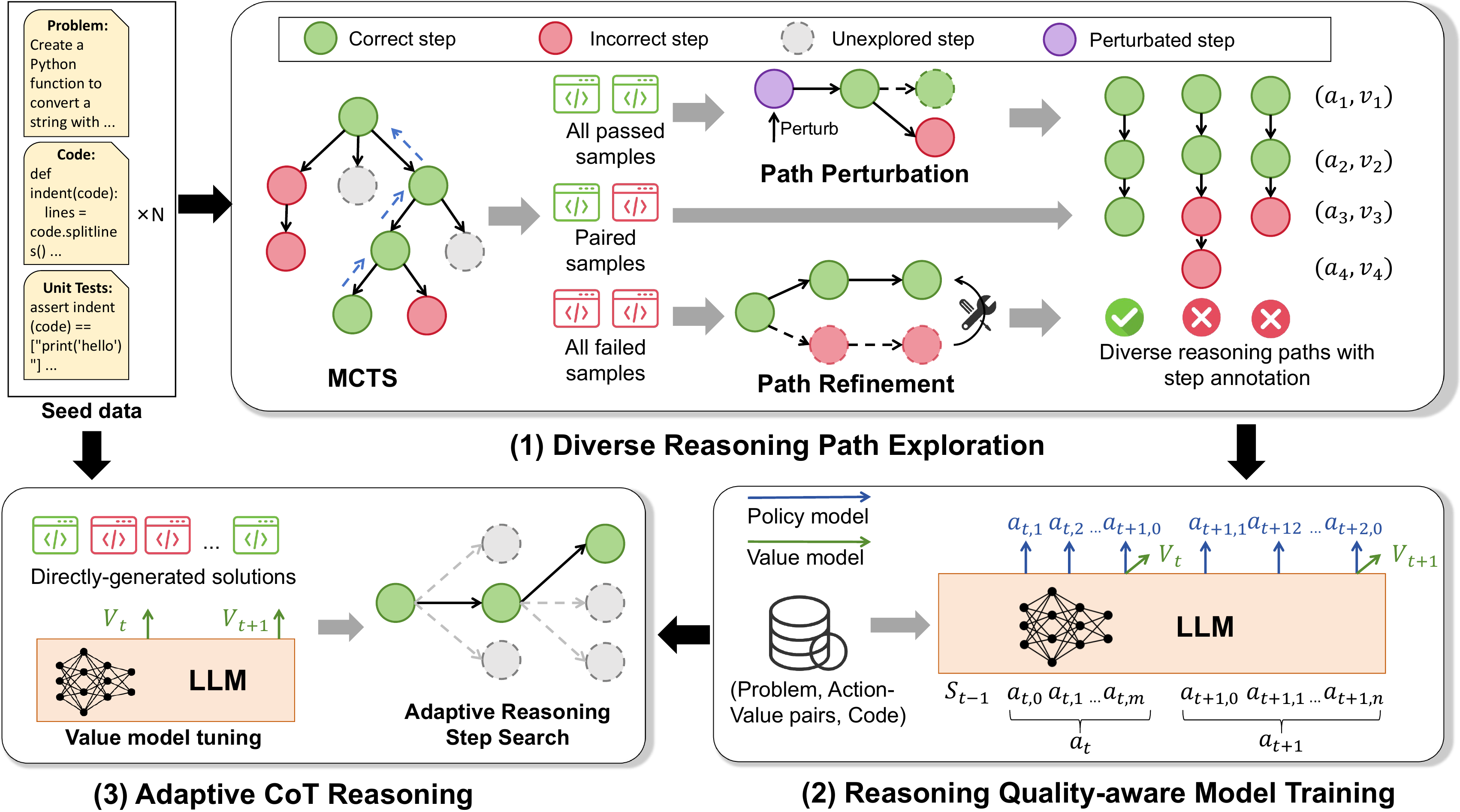}
    \caption{The overall framework of \tool.}
    \label{fig:overall}
    \vspace{-0.2cm}
\end{figure*}

The overall procedure of \tool is illustrated in Figure~\ref{fig:overall}. The framework consists of three main phases. In the \textbf{Diverse reasoning path exploration} phase, we collect high-quality {and} diverse reasoning data with detailed annotations of reasoning step quality, for {subsequent} model training. We start with a seed dataset $D_{seed} = {(q^i, c^i, u^i)}^N_{i=1}$ containing programming problems, code solutions, and test cases. For each sample, we first employ MCTS to construct reasoning trees through iterative selection, expansion, evaluation, and backpropagation. We then {propose to} enhance dataset diversity through \textit{path perturbation} for unexplored incorrect solutions and \textit{path refinement} for unexplored correct solutions. Based on the self-explored data, in the \textbf{Reasoning quality-aware model training} phase, we develop two models: a policy model trained on correct reasoning paths through reinforced fine-tuning to produce effective reasoning, and a value model trained on both correct and incorrect reasoning paths to assess reasoning step quality. Finally, in the \textbf{Adaptive CoT reasoning} phase, we teach the model to dynamically {switch}
between direct generation and step-by-step reasoning for different problems. {Specifically, we design {an adaptive reasoning step search algorithm,}
where the value model is continually fine-tuned and collaborates with the policy model to choose the most suitable reasoning approach.}

\subsection{Diverse Reasoning Path Exploration}
This part aims at collecting high-quality and diverse reasoning data with detailed annotations of reasoning step quality for model training. We first perform customized MCTS for initial data collection, followed by path perturbation and refinement to further improve {the diversity of reasoning data.}

\subsubsection{Monte Carlo Tree Search {(MCTS)}}
As shown in Figure~\ref{fig:mcts}, during MCTS, \tool starts with the initial state as its root and systematically grows the search tree by adding new nodes. Specifically, within the context of code generation, we customize the four key operations of the MCTS algorithm as follows:

\textbf{Selection.} During each iteration of the MCTS, the selection process begins with $\mathbf{s}_0$, representing the initial state containing the input problem $q$. The algorithm then proceeds to explore the tree $\mathcal{T}_k$ and selects the children node with the highest PUCT score~\cite{rosin2011multi}. The PUCT score incorporates the log-probability from pre-trained LLM as prior knowledge to rank each node and balance exploration and exploitation during sampling based on visiting history. This selection process is mathematically represented as:
\begin{equation}
\mathbf{a}_t = \arg\max_{\mathbf{a}\in \mathcal{T}_k}[ \hat{Q}(\mathbf{s},\mathbf{a}) + c_{\text{puct}}\pi_{\theta_k}(\mathbf{a}|\mathbf{s}_t)\sqrt{\frac{N_{\text{parent}}(\mathbf{a})}{1+N(\mathbf{s}_t,\mathbf{a})}} ]
\end{equation}
where the state-action value $\hat{Q}(\mathbf{s},\mathbf{a})$ and its visiting count $N(\mathbf{s},\mathbf{a})$ are initialized to 0 and will be updated in the Backpropagation stage. $N_{\text{parent}}(\mathbf{a})$ represents the visiting count of the parent node of $\mathbf{a}$. The prior $\pi(\mathbf{a}|\mathbf{s}_t)$ is the exponential of the averaged log-probability of all tokens in the step $\mathbf{a}$, i.e., $\exp\left(\frac{1}{|\mathbf{a}|}\sum\log\pi(\mathbf{a}_j|\mathbf{a}_{<j},\mathbf{s}_t)\right)$. $c_{\text{puct}}$ is a weight control parameter and we use its default value 0.5. Therefore, the node with high log-probability and low visiting counts will be prioritized. The action selection iterates until it encounters a leaf node of the current search tree. 

\textbf{Expansion.} In this stage, we expand the node by generating the next step. Specifically, we first trace the selected leaf node to the root and obtain a partial solution $\mathbf{s}_t=\{\mathbf{s}_0, \mathbf{a}_1, ... \mathbf{a}_t\}$. The partial solution contains the problem description and current generated steps and serves for further node expansion. Different from traditional reinforcement learning scenario where actions are always limited~\cite{silver2016mastering}, given that the LLM can theoretically generate a nearly unlimited number of potential reasoning steps~\cite{guo2025deepseek}, we employ sampling-based generation to obtain promising and diverse steps. 
To accurately identify the end of each step generation, we utilize in-context learning with XML-style like delimiters ``$\langle\text{step}\rangle$'' and ``$\langle\backslash\text{step}\rangle$'' to wrap each step. The LLM output is truncated after the closing delimiter ``$\langle\backslash\text{step}\rangle$''.

\textbf{Evaluation.} In this stage, we evaluate the quality of the newly expanded node $\mathbf{s}_t$. Specifically, if it reaches a leaf node that contains the final solution code, we evaluate its correctness using test cases $u$ and compute its reward according to the following function:  

\begin{equation}
\hat{V}(\mathbf{s}_t)  = r  =
\begin{cases} 
\text{1}, & \text{if all test cases pass} \\ 
\text{-1}, & \text{if any test case fails}
\end{cases}
\end{equation}

For non-terminal expanded nodes, we set the value to 0 since there are no clear signals indicating their correctness at intermediate stages.

\textbf{Backpropagation.}  At the end of each iteration, we update each node $(\mathbf{s})$ along the path from the leaf node $\mathbf{s}_t$ to the root $\mathbf{s}_0$. Following~\cite{silver2016mastering,DBLP:journals/corr/abs-2405-03553}, we update their visiting counts and state-action values according to the following rules:
$N(\mathbf{s},\mathbf{a}) \leftarrow N(\mathbf{s},\mathbf{a}) + 1$ and $\hat{Q}(\mathbf{s},\mathbf{a}) \leftarrow \frac{1}{N(\mathbf{s},\mathbf{a})}\sum_{j=1}^N\mathbb{I}_{\mathbf{s},\mathbf{a}\rightarrow\mathbf{s}_t}\hat{V}(\mathbf{s}_t)^{(j)}$. The $\hat{Q}(\mathbf{s},\mathbf{a})$ measures the probability of generating
correct code when taking action $a$ at state $s$, i.e., the value represents the probability of generating a correct solution when continuing from the current path.

The above four stages will continue until reach the maximum iteration numbers. After that we obtain the final tree $\mathcal{T}$ for each problem, which stores the expanded nodes and their corresponding state-action values $Q(\mathbf{s},\mathbf{a})$. 


\subsubsection{Path Perturbation and Refinement}
Despite leveraging MCTS, the constructed trees do not always include both correct and incorrect paths for all programming problems. For example, for DeepSeek-Coder-6.7B-Instruct, 29.5\% samples during the collection process contain only correct paths as the problems might be too simple for the model to solve, while 11.8\% samples contain only incorrect paths due to the model's limited capabilities. The lack of correct reasoning training data can limit the policy model's generation capabilities, while a lack of incorrect paths can impair the value model's ability to assess the quality of intermediate steps. To address this issue, we propose path perturbation and refinement methods to enrich the {reasoning}
data for problems that include only {correct or incorrect}
paths, with details as below.

(1) \textit{For samples including only correct reasoning paths}, we propose path perturbation to create incorrect paths for them. \gsz{This approach improves the model's quality assessment ability for basic programming problems where MCTS fails to identify incorrect reasoning paths. Specifically, we first randomly partition the original correct reasoning path into two segments: ${\mathbf{s}_0, \mathbf{a}_1, ... \mathbf{a}s}$ and ${\mathbf{a}{s+1}, ... \mathbf{a}_t}$. Next, we retain only the first segment while perturbing the problem description at the root node by preserving just the initial sentence. Given this ambiguous requirement and the partial reasoning steps, LLMs generate subsequent steps and code that do not well fit the original requirement. Finally, after verifying through test cases that the generated code is incorrect, we replace the ambiguous requirement with the original specification to create an incorrect reasoning path. We further provide an example in our replicate package for illustration~\cite{replicate}. This process simulates scenarios where the model hallucinates or misinterprets requirements, which is a common occurrence in LLM-generated code~\cite{DBLP:journals/corr/abs-2409-20550}.}


(2) \textit{For samples including only incorrect reasoning paths}, we propose path refinement to synthesize correct reasoning paths for them. Specifically, we propose to leverage the ground truth solutions to refine the reasoning process through self-reflection~\cite{DBLP:conf/nips/MadaanTGHGW0DPY23}. As shown in Figure~\ref{fig:overall}, using a reflection prompt template (available in
the GitHub repository~\cite{replicate}), we present the failed reasoning paths alongside the ground truth solution, guiding the model to identify specific errors and generate refined reasoning paths.

After these steps, we construct our reasoning dataset for training $D_{train}$ by sampling paths from the MCTS trees $\mathcal{T}$, along with new paths generated through path perturbation and refinement processes. For the actions,
we {calculate their}
values differently based on the path source. For paths from MCTS and path perturbation, we obtain values from their respective MCTS trees. Specifically, according to the transition function $Q(\mathbf{s}_t,\mathbf{a}_t) = r(\mathbf{s}_t,\mathbf{a}_t) + V(\mathbf{s}_{t}) = V(\mathbf{s}_{t+1})$ for non-terminal nodes~\cite{DBLP:journals/corr/abs-2405-03553,silver2016mastering}, we could employ the $Q(\mathbf{s}_t,\mathbf{a}_t)$ as the value $\mathbf{v}_t$ of action $\mathbf{a}_t$.
For paths obtained through path refinement, we assign a value of 1 to the final state. Finally, each sample in $D_{train}$ consists of a problem description $q$ paired with a sequence of reasoning steps ${(\mathbf{a}_1,\mathbf{v}_1),...(\mathbf{a}_n,\mathbf{v}_n)}$, where $\mathbf{a}_i$ and $\mathbf{v}_i$ represent the content of the step and its associated value, respectively.

\subsection{Reasoning Quality-aware Model Training}\label{sec:dual}
This part aims at training LLMs to generate high-quality reasoning steps and select the best generated step. Following the practice in AlphaZero~\cite{silver2017mastering}, we propose a dual reasoning system that contains a policy model and a value model. We employ the pre-trained LLM as the policy model $\pi_{\theta}$. \gsz{In this work, we specifically use fine-tuned DeepSeek-Coder-6.7B-Instruct and Qwen2.5-Coder-7B-Instruct as our policy models}. To save cost, we do not introduce another model as the value model $V_\phi$ but extend the policy model by only adding an auxiliary linear layer with the tanh activation function~\cite{lecun2015deep}, which works alongside the traditional softmax layer responsible for token prediction, as shown in Figure~\ref{fig:overall} {(2)}. This design implies that these two models, $\pi_\theta$ and $V_\theta$, share most of their parameters. Based on the correct paths $X^+$ and incorrect paths $X^-$ from our constructed dataset $D_{train}$, we train the policy and value models through multi-task learning as follows:

\begin{equation}\label{eq:1}
\displaystyle \arg\min_{\theta,\phi} \sum_{\mathbf{x} \in X^+} -\log \pi_\theta(\mathbf{x}|\mathbf{q}) + \beta\sum_{\mathbf{x} \in X^+ \cup X^-}\sum_{t=1}^{T(\mathbf{x})}|V_\phi(\mathbf{s}_t) - \mathbf{v}_t|^2
\end{equation}
where $\mathbf{x} = \{\mathbf{a}_1,\mathbf{a}_1,...,\mathbf{a}_n\}$. The first term represents the negative log-likelihood loss for reinforced fine-tuning the policy model on correct paths, and the second term captures the loss in value prediction for both correct and incorrect solutions, respectively. $T(\mathbf{x})$ denotes the number of steps for path $\mathbf{x}$. $\beta$ is a tunable hyper-parameter to control the weight of value model loss.

\begin{algorithm}[t]
\caption{Adaptive reasoning step search algorithm}
\label{algo:sbs}
\begin{algorithmic}[1]
\REQUIRE Beam sizes $B_1$, $B_2$, question $\mathbf{s}_0$, max steps $T$.\\
\ENSURE Top-1 results\\
\STATE $C \gets [\mathbf{s}_0] \times B_1$, $t \gets 0$ 

\WHILE{$t < T$ \textbf{and} non-terminal path in $C$}
\STATE Initialize priority queue $C_{t+1}$ 

\FOR{$\mathbf{s}_t \in C$}
\IF{t==0}
\STATE Sample $\{a^{(b)}\}_{b=1}^{\lceil B_2 /2 \rceil} \sim \pi_\theta(a \mid \mathbf{s}_t)$
\STATE Sample $\{a^{(b)}\}_{b=1}^{\lfloor B_2/2 \rfloor} \sim \pi_\theta(a \mid \mathbf{s}_t, \texttt{\`{}\`{}\`{}\{lang\}})$
\ELSE
\STATE Sample $\{a^{(b)}\}_{b=1}^{B_2} \sim \pi_\theta(a \mid \mathbf{s}_t)$ 
\ENDIF
\ENDFOR

\FOR{$b = 1$ to $B_2$}
\STATE $\mathbf{s}_{t+1} \gets \text{Concat}\big[\mathbf{s}_t, a^{(b)}\big]$
\STATE Add $\big(\mathbf{s}_{t+1}, V_\phi(\mathbf{s}_{t+1})\big)$ to $C_{t+1}$
\ENDFOR

\STATE $C \gets \text{Top-}B_1 \text{ of } C_{t+1}$
\ENDWHILE

\RETURN Top-1 of $C$ 
\end{algorithmic}
\end{algorithm}

\subsection{Adaptive CoT Reasoning}
This stage aims at teaching models to apply chain-of-thought reasoning adaptively. Based on the step evaluation capability of the value model, we propose extending it to judge whether to use CoT reasoning for different problems. To this end, we first collect training data containing directly-generated solutions and their value scores by the models themselves,
then continually fine-tune the value model on this data. 
Specifically, as shown in Figure~\ref{fig:overall} (3), we feed the LLM with the seed data $D_{seed}$
and force the model to directly output the code without CoT reasoning. We then evaluate these directly-generated outputs using test cases $u$. By assigning the value of correct solutions as 1 and incorrect solutions as -1, we build the adaptive reasoning dataset $D_{ar}$ that contains the question, code, and value tuple $(q, a, v)$. To prevent catastrophic forgetting during continual training~\cite{DBLP:conf/icse/GaoZGW23,mccloskey1989catastrophic}, 
we incorporate a KL divergence regularization term on the policy model's predictions. In addition, to maintain balance between correct and incorrect samples, we perform downsampling for the combined dataset, i.e., $D_{train}$ and $D_{ar}$, and train the model as follows: 

\begin{equation}\label{eq:2} 
\arg\min_{\theta,\phi} \sum_{\mathbf{x} \in X^+} KL( \pi_\theta(\mathbf{x}|\mathbf{q}), \pi_{old}(\mathbf{x}|\mathbf{q})) + \beta\sum_{\mathbf{x} \in X^+ \cup X^-}\sum_{t=1}^{T(\mathbf{x})}|V_\phi(\mathbf{s}_t) - \mathbf{v}_t|^2
\end{equation}
where $X^+$ and $X^-$ denotes the correct paths and incorrect paths from the concatenation dataset, respectively. $\pi_{old}$ is the model obtained in Section~\ref{sec:dual}.


During model inference, we
propose {an}
adaptive reasoning step search algorithm {by employing the trained policy and value model}, as detailed in Algorithm~\ref{algo:sbs}. 
{The overall procedure follows traditional beam search while using the value model to guide {the decoding process.}
It contains two beam size parameters: $B_1$ which represents the number of preserved current best samples (similar to traditional beam search), and $B_2$ which controls the number of candidate actions generated by the policy model at each step.} 
{The process operates as follows: In the first iteration ($t=0$), the policy model generates $\lceil B_2 /2\rceil$ reasoning steps based on the programming problem $\mathbf{s}_0$, and $\lfloor B_2 /2\rfloor$ solutions without reasoning. To force the model to {generate solutions without reasoning,}
the response prefix $\texttt{\`{}\`{}\`{}\{lang\}}$ is appended to the end of the prompt, where \texttt{lang} is the target programming language (lines 5–7). Then, the algorithm leverages the value predictions from the value model $V_\phi$ to rank the steps and directly-generated solution, determining whether to use
step-by-step reasoning {or not}. The newly expanded states $\mathbf{s}_{t+1}$ and predicted value are stored in a priority queue $C_{t+1}$ to enable efficient beam search (lines 12–14). From the second iteration onward, the policy model generates $B_2$ reasoning steps based on current paths in $C$ (lines 8-9) and the value model assesses the quality of each reasoning path as in the first iteration. Finally, this process continues until either the maximum number of steps $T$ is reached or all paths are terminated. At the end of the search, the top candidate is returned as the final answer (line 18).}

\section{EXPERIMENTAL setup}\label{sec:setup}

\subsection{Research Questions}
In the evaluation, we focus on the following four research questions:

\begin{enumerate}[label=\bfseries RQ\arabic*:,leftmargin=.5in]
    \item How effective is \tool in improving chain-of-thought code generation?
    \item What are the contributions of different modules in \tool?
    \item What is the impact of different hyper-parameters on the performance of \tool?
\end{enumerate}

To study RQ1, we conduct a comprehensive evaluation of \tool on three widely used code generation benchmarks and compare it with four representative baseline methods across two state-of-the-art open-source code LLMs. 
For RQ2, we remove different parts in \tool to assess their individual contributions. For RQ3, we perform a detailed analysis of hyperparameters in \tool. 

\subsection{Benchmarks and Metrics} Following previous work \cite{DBLP:conf/icml/0003W0D024, wang2024exploring}, we evaluate performance on three widely-used coding benchmarks: MBPP~\cite{austin2021program}, HumanEval~\cite{chen2021evaluating}, and LiveCodeBench~\cite{jain2024livecodebench}. 

\begin{itemize}
\item MBPP~\cite{austin2021program} consists of 378 crowd-sourced Python programming problems, which is hand-verified by the original authors. We follow previous work \cite{wang2024exploring} and report results on both the original and plus versions, which include more test cases ~\cite{DBLP:conf/nips/LiuXW023}. 
\item HumanEval~\cite{chen2021evaluating} consists of 164 manually-written Python programming problems, each accompanied by an average of 9.6 test cases. Following previous work, we report the performance of both original and plus versions ~\cite{DBLP:conf/nips/LiuXW023}. 
\item LiveCodeBench (LCB)~\cite{jain2024livecodebench} is a contamination-free programming benchmark that collects new problems over time from contest platforms. We include all available 713 programming problems during the experimental period. 
\end{itemize}

To measure the accuracy of the predicted code, we use the widely-adopted pass rate metric (pass@1) \cite{wang2024exploring, DBLP:conf/iclr/LuoX0SGHT0LJ24,DBLP:conf/icml/0003W0D024}.


\begin{table*}[t]
\centering
\caption{Comparison results with baseline models. The \textbf{bold} figures indicate the best results.}
\scalebox{0.85}{
\begin{tabular}{ll|c|cc|cc|c}
\midrule
\multicolumn{2}{c|}{\multirow{2}{*}{Approach}} & \multirow{2}{*}{Size/Version} & \multicolumn{2}{c|}{MBPP} & \multicolumn{2}{c|}{HumanEval} & LiveCodeBench \\
& & & Original & Plus & Original & Plus & Avg \\
\midrule
\multicolumn{8}{c}{Closed-source Proprietary Models} \\
\midrule
\multicolumn{2}{c|}{GPT-4o} & Nov. 2024 & 87.0 & 71.4 & 92.1 & 87.2 & 45.0 \\
\multicolumn{2}{c|}{GPT-4o-mini} & July. 2024 & 85.4 & 72.2 & 88.4 & 83.5 & 39.4 \\
\multicolumn{2}{c|}{GPT-4-Turbo} & Apr. 2024 & 85.7 & 73.3 & 90.2 & 86.6 & 44.2 \\
\multicolumn{2}{c|}{Gemini 1.5 Flash 002} & Sept. 2024 & 68.0 & 59.0 & 82.3 & 75.6 & 36.2 \\
\midrule
\multicolumn{8}{c}{15B+ Models} \\
\midrule
\multicolumn{2}{c|}{DeepSeek-Coder-V2-Instruct} & 236B & 89.4 & 75.1 & 85.4 & 82.3 & 41.9 \\
\multicolumn{2}{c|}{Mixtral-8x22B-Instruct-v0.1} & 176B & 73.8 & 64.3 & 76.2 & 72.0 & 24.2 \\
\multicolumn{2}{c|}{Llama3-70B-Instruct} & 70B & 82.3 & 69.0 & 77.4 & 72.0 & 26.2 \\
\multicolumn{2}{c|}{DeepSeek-Coder-33B-Instruct} & 33B & 80.4 & 70.1 & 81.1 & 75.0 & 23.4 \\
\multicolumn{2}{c|}{StarCoder2-15B-Instruct-v0.1} & 15B & 75.2 & 61.2 & 67.7 & 60.4 & 14.6 \\
\midrule
\multicolumn{8}{c}{7B-level Models} \\
\midrule
\multirow{6}{*}{DeepSeek-Coder-6.7B-Instruct} & Direct Generation & 6.7B & 72.7 & 63.4 & 73.8 & 70.1 & 18.5 \\
& Self Planning & 6.7B & 74.1 & 62.4 & 77.4 & 71.3 & 18.7 \\
& Code CoT & 6.7B & 73.8 & 60.5 & 74.4 & 65.9 & 18.2 \\
& SFT & 6.7B & 72.2 & 61.9 & 73.2 & 65.9 & 18.1 \\
& DS-R1's RL & 6.7B & 74.3 & 65.1 & 77.4 & 71.3 & 20.1 \\
& \tool & 6.7B & \textbf{82.0} & \textbf{72.2} & \textbf{82.9} & \textbf{73.8} & \textbf{23.8} \\
\midrule
\multirow{6}{*}{Qwen2.5-Coder-7B-Instruct} & Direct Generation & 7B & 83.5 & 71.7 & 88.4 & 84.1 & 33.2 \\
& Self Planning & 7B & 84.1 & 72.0 & 84.1 & 78.0 & 29.2 \\
& Code CoT & 7B & 81.2 & 69.6 & 81.1 & 74.4 & 28.1 \\
& SFT & 7B & 82.0 & 71.7 & 86.0 & 81.1 & 28.1 \\
& DS-R1's RL & 7B & 84.7 & 71.7 & 89.0 & 84.8 & 34.1 \\
& \tool & 7B & \textbf{88.6} & \textbf{75.1} & \textbf{90.9} & \textbf{86.0} & \textbf{36.7} \\
\bottomrule
\end{tabular}
}
\label{tab:results}
\end{table*}

\subsection{Models and Baselines} In this paper, we select two popular and state-of-the-art code LLMs for experiments:

\begin{itemize}
    \item \textbf{DeepSeek-Coder} (DSC) is trained from 2T tokens from scratch, achieving state-of-the-art performance on a variety of code intelligence tasks. Specifically, we choose DeepSeek-Coder-6.7B-Instruct in this paper.
    \item \textbf{Qwen2.5-Coder} (QC) is a family of LLMs based on models of the Qwen2.5 series and continues to be pretrained on a vast corpus of over 5.5 trillion tokens, presenting state-of-the-art performance in both code generation and reasoning. We choose the 7B version of Qwen2.5-Coder-Instruct.
\end{itemize}

\gsz{We compare our method with recent representative COT-based code generation methods from top venues, including two representative prompting-based COT code generation methods, Self-planning and Code COT, and a training-based SFT method used in~\cite{DBLP:journals/tse/YangZCZZC24}. Besides, considering the popularity of DeepSeek-R1, we further include its training method as a baseline.}
 
\begin{itemize}
    \item \textbf{Self-Planning}~\cite{DBLP:journals/tosem/JiangDWFSLJJ24} first leverage LLMs to plan concise solution steps from the intent combined with a few-shot prompt and then generate code step by step, guided by the preceding solution steps. We replicate it based on the few-shot examples provided in their paper.
    \item \textbf{Code CoT}~\cite{DBLP:journals/tosem/JiangDWFSLJJ24} is a prompting method that asks LLMs to describe the implementation code line by line and then write the code.  We also replicate it based on the prompt provided in the paper.
    \item \textbf{SFT}~\cite{DBLP:journals/tse/YangZCZZC24} fine-tunes LLMs with data that contain chain-of-thought prompts and final solution code. Specifically, we leverage the data provided by COTTON~\cite{DBLP:journals/tse/YangZCZZC24} which is crafted by distilling CoTs from GPT-4.
    \item \textbf{DS-R1's RL}~\cite{guo2025deepseek} is a reinforcement learning method used to train DeepSeek-R1-Zero. It is the state-of-the-art LLM reasoning enhancement method. We replicate it based on the code provided by the Open-R1 repository~\cite{open-r1}. We train the model with the same seed data used in \tool and employ the same reward functions as used in DeepSeek-R1-Zero, namely the format reward and the accuracy reward.
\end{itemize}
 
 We also compare \tool with other open-source models with much larger parameters such as StarCoder2-15B-Instruct-v0.1~\cite{DBLP:journals/corr/abs-2402-19173}, DeepSeek-Coder-33B-Instruct~\cite{guo2024deepseek}, Llama3-70B-Instruct~\cite{DBLP:journals/corr/abs-2407-21783}, Mixtral-8x22B-Instruct-v0.1~\cite{DBLP:journals/corr/abs-2401-04088}, and DeepSeek-Coder-V2-Instruct~\cite{zhu2024deepseek}. We also compare with some closed-source proprietary models including \gsz{GPT-4o, GPT-4o-mini,} 
 GPT-4~\cite{DBLP:journals/corr/abs-2303-08774} and Gemini 1.5 Flash 002~\cite{DBLP:journals/corr/abs-2312-11805}.

\subsection{Implementation details} 
\gsz{We employ Self-OSS~\cite{DBLP:conf/nips/0003C0DJMVWG024} as our seed dataset for two key reasons: first, its correctness can be reliably assessed through unit tests, and second, the authors of Self-OSS have conducted a thorough data decontamination process~\cite{DBLP:conf/nips/0003C0DJMVWG024} to remove questions that overlap with popular benchmarks}. To save experimental costs, we 
randomly select 10K question-code pairs from Self-OSS for MCTS data collection. During the expansion stage of MCTS, we employ in-context learning and incorporate three demonstrations which are randomly selected from a pool of eight prepared examples. These eight examples align with previous few-shot CoT code generation methods~\cite{DBLP:journals/tosem/JiangDWFSLJJ24}. The detailed prompt templates used in this paper can be found in the GitHub repository~\cite{replicate}. \gsz{We construct one search tree per sample in the seed data, with an average of 254 and 261 nodes per tree for Deepseek-Coder and Qwen2.5-coder, respectively. When extracting paths from MCTS tree to construct $D_{train}$, we first filter paths with incorrect formats and randomly sample up to three correct and three incorrect reasoning path for each problem. We then perform downsampling to keep a balance between positive and negative samples. This yields finetuning datasets of 68K and 60K samples, respectively. Similarly, we also conduct the same data sampling process for data generated in the path perturbation and refinement process. The initial data construction takes 43 and 47 hours respectively, but this process is performed only once and the data can be reused.}

Following previous studies~\cite{DBLP:conf/icml/0003W0D024}, we train the model for two epochs with a batch size of 128. During inference, we set the value loss weight $\beta$ to 0.025, the LLM temperature to 0.5, and configure $B_1$ and $B_2$ to 3 and 4, respectively. The influence of these parameters is discussed in Section~\ref{sec:para}. \gsz{Finetuning with data from MCTS requires approximately 12 hours and 10 hours on one GPU for these two models, which is similar to previous work~\cite{DBLP:conf/icml/0003W0D024} when using the same training data size and epochs.} We repeat \tool three times and report its average results to eliminate the influence of sampling and fluctuations in LLM. During adaptive CoT reasoning training, incorporating KL divergence increases GPU memory requirements. Due to our server's memory limitations, we employ the parameter-efficient tuning~\cite{DBLP:journals/pacmse/WangFGGLPHDL25,han2024parameter} method LoRA~\cite{DBLP:conf/iclr/HuSWALWWC22} to prevent out-of-memory errors. For Qwen, we observed that the model already generates high-quality Chain-of-Thought reasoning with prompting alone, and additional fine-tuning might degrade performance. Therefore, we apply KL divergence during the first training stage for Qwen as well. All experiments are conducted on a server running Ubuntu 20.04 with 4 NVIDIA Tesla A100 GPUs (40GB memory each).

\section{Experiments}\label{sec:result}

\subsection{RQ1: Performance Evaluation}
To evaluate the effectiveness of \tool in improving the reasoning capabilities of code LLMs, 
we compare it with 
multiple baselines
across three different evaluation benchmarks. 
Table~\ref{tab:results} presents the performance of \tool and baseline methods on MBPP, HumanEval, and LiveCodeBench. Based on these results, we derive the following findings:

\textbf{Simple prompting and SFT provide limited improvements in LLMs' reasoning ability for code generation.}
As shown in Table~\ref{tab:results}, we observe that simple prompting and SFT have limited effectiveness in enhancing model reasoning capabilities. For instance, DeepSeek-Coder-6.7B-Instruct with Self-Planning prompt shows only a marginal improvement on HumanEval Original from 73.8\% to 77.4\%, while experiencing a decrease on MBPP Plus from 63.4\% to 62.4\%. Similarly, Qwen2.5-Coder-7B-Instruct with SFT exhibits a reduction in performance on the MBPP Original benchmark, declining from 83.5\% to 82.0\%
compared to the base model. 
These results suggest that merely applying prompt engineering or supervised fine-tuning fails to effectively improve reasoning ability for complex code generation tasks.

\textbf{\tool effectively improve models' reasoning capabilities in code generation.}
As shown in Table~\ref{tab:results}, models augmented with \tool demonstrate substantial improvements across all evaluation benchmarks. For DeepSeek-Coder-6.7B-Instruct, using \tool improves performance by 9.3\%, 9.1\% and 5.3\% in terms of pass@1 on MBPP Original, HumanEval Original and LiveCodeBench, respectively. Notably, \tool outperforms the state-of-the-art baseline DS-R1's RL by 7.7\%, 5.5\%, and 3.7\% on these same benchmarks. Similarly, Qwen2.5-Coder-7B-Instruct enhanced with \tool shows remarkable gains, improving by 5.1\%, 2.5\% and 3.5\% on MBPP Original, HumanEval Original and LiveCodeBench compared to the base model, while surpassing DS-R1's RL by 3.9\%, 1.9\%, and 2.6\% respectively. These consistent improvements across multiple benchmarks and the clear advantage over other reasoning enhancement techniques like DS-R1's RL demonstrate \tool's effectiveness in augmenting model reasoning capabilities for code generation tasks.

\textbf{Small models augmented with \tool 
outperform some larger models.}
By comparing the performance of models of different sizes, we find that smaller models enhanced with \tool can outperform certain larger models. For example, Qwen2.5-Coder-7B-Instruct+\tool achieves 88.6\% and 75.1\% on MBPP Original and Plus, \gsz{outperforming GPT-4o, which scores 87.0\% and 71.4\%, respectively}.
Similarly, DeepSeek-Coder-6.7B-Instruct+\tool attains 82.9\% on HumanEval Original, exceeding both Llama3-70B-Instruct (77.4\%) and DeepSeek-Coder-33B-Instruct (81.1\%) with 4-10 times more parameters. 
These results highlight that through \tool, smaller models can achieve comparable or superior performance to larger models in code generation.

\begin{tcolorbox}[width=\linewidth,boxrule=0pt,top=1pt, bottom=1pt, left=1pt,right=1pt, colback=gray!20,colframe=gray!20]
\textbf{Answer to RQ1:} 
\tool is highly effective in improving CoT-based code generation, consistently improving performance across all model sizes and enabling smaller models to outperform much larger ones.
\end{tcolorbox}

\subsection{RQ2: Ablation Study}\label{sec:ablation}

\subsubsection{Impact of different components} 
In this section, we validate the contribution of different components of \tool, including the path perturbation and refinement in diverse reasoning path exploration, the value model-based reasoning quality assessment, and adaptive CoT reasoning.

\textbf{Path perturbation and path refinement.} To assess the impact of our path exploration strategies, we independently remove them from the path exploration process. Path refinement augments 3,051 and 3,499 correct reasoning paths for \gsz{DeepSeek-Coder} and \gsz{Qwen2.5-Coder}, respectively, while path perturbation provides 7,834 and 6,169 negative training samples for these models. As shown in Table~\ref{tab:ablation}, 
\gsz{removing path perturbation leads to obvious performance degradation on two relatively basic datasets MBPP Plus and HumanEval Plus. For example, Deepseek-Coder's performance on HumanEval Plus drops from 73.8\% to 70.7\% when this component is excluded. This shows that path perturbation could enhance the model's quality assessment on basic programming problems}. Similarly, removing path refinement leads to a larger degradation, particularly on MBPP Plus where \gsz{DeepSeek-Coder} experiences a \gsz{3.1\%} 
decrease. 
decrease. 
These results demonstrate that both path exploration strategies play crucial roles in improving model generalization across different scenarios by enhancing the diversity of collected paths. 

\textbf{Value model-based reasoning quality assessment.} We further investigate the importance of the value model-based reasoning quality assessment 
by removing the validation of intermediate reasoning steps, i.e., changing $V_\phi(\mathbf{s}_{t+1})$ in line 14 in Algorithm~\ref{algo:sbs} to 0 and selecting $C$ randomly in line 16. This leads to tremendous performance drops, with DeepSeek's performance on LiveCodeBench decreasing by \gsz{5.2\% } 
and Qwen2.5's by \gsz{5.4\%.} 
This demonstrates the critical importance of validating the quality of intermediate reasoning steps during inference. We provide a case in Section~\ref{sec:case} for further illustration.

\begin{table}[t]
\centering
\caption{\gsz{Ablation study of different components in \tool. The \textbf{bold} figures indicate the best results, where PP and PR denote path perturbation and path refinement, respectively.}}
\scalebox{0.83}{
\begin{tabular}{l|cc|cc|c}
\midrule
\multirow{2}{*}{Approach}  & \multicolumn{2}{c|}{MBPP} & \multicolumn{2}{c|}{HumanEval} & LCB \\
& Original & Plus & Original & Plus & Avg \\
\midrule
\multicolumn{6}{c}{DeepSeek-Coder-6.7B-Instruct} \\
\midrule
\tool  & \textbf{82.0±1.1} & \textbf{72.2±0.6} & \textbf{82.9±1.0} & 73.8±1.2 & \textbf{23.8±0.5} \\
-w/o PP & 81.2±0.4 & 70.2±1.9 & 80.7±1.8 & 70.7±0.6 & 23.5±0.3 \\
-w/o PR & 80.9±0.9 & 69.1±1.2 & 80.3±1.5 & \textbf{74.6±0.7} & 23.3±0.6 \\
-w/o value model & 76.6±0.8 & 64.6±0.3 & 77.3±1.3 & 69.9±1.4 & 18.6±0.4 \\
-w/o adaptive cot & 81.0±0.9 & 69.0±1.7 & 79.3±1.1 & 70.3±2.2 & 21.9±1.4 \\
\midrule
\multicolumn{6}{c}{Qwen2.5-Coder-7B-Instruct} \\
\midrule
\tool & \textbf{88.6±0.9} & \textbf{75.1±0.3} & \textbf{90.9±0.0} & \textbf{86.0±2.2} & 36.7±0.3 \\
-w/o PP & 87.3±1.0 & 73.5±1.1 & 89.8±0.3 & 84.6±1.0 & \textbf{37.0±0.4} \\
-w/o PR & 87.0±0.7 & 73.6±0.6 & 90.0±0.8 & 85.8±1.2 & 35.9±0.4 \\
-w/o value model & 84.1±0.9 & 71.6±0.5 & 88.8±0.3 & 83.3±1.3 & 31.3±0.6 \\
-w/o adaptive cot & 86.4±0.8 & 73.4±1.2 & 87.7±0.8 & 83.9±0.8 & 35.2±0.6 \\
\midrule
\end{tabular}
}
\label{tab:ablation}
\vspace{-0.3cm}
\end{table}

\begin{table}[t]
\centering
\caption{\gsz{The impact of different training strategies. The \textbf{bold} figures indicate the best results.}}
\scalebox{0.82}{
\begin{tabular}{l|cc|cc|c}
\midrule
\multirow{2}{*}{Approach}  & \multicolumn{2}{c|}{MBPP} & \multicolumn{2}{c|}{HumanEval} & LCB \\
& Original & Plus & Original & Plus & Avg \\
\midrule
\multicolumn{6}{c}{DeepSeek-Coder-6.7B-Instruct} \\
\midrule
\tool  & \textbf{82.0±1.1} & \textbf{72.2±0.6} & \textbf{82.9±1.0} & \textbf{73.8±1.2} & \textbf{23.8±0.5} \\
Only first stage & 81.0±2.0 & 70.4±1.8 & 80.4±0.4 & 71.8±2.5 & 18.7±0.9 \\
Two stage no KL & 77.8±0.6 & 67.5±0.3 & 76.0±2.1 & 67.6±0.8 & 19.5±1.1 \\
\midrule
\multicolumn{6}{c}{Qwen2.5-Coder-7B-Instruct} \\
\midrule
\tool & \textbf{88.6±0.9} & \textbf{75.1±0.3} & \textbf{90.9±0.0} & \textbf{86.0±2.2} & \textbf{36.7±0.3} \\
Only first stage & 86.8±1.6 & 73.4±1.2 & 88.0±0.9 & 83.3±1.3 & 34.4±0.2 \\
Two stage no KL & 74.3±0.7 & 64.5±0.9 & 72.2±0.3 & 67.1±1.2 & 23.6±0.5 \\
\midrule
\end{tabular}
}
\label{tab:ablation2}
\vspace{-0.5cm}
\end{table}

\begin{figure*}[t]
    \centering
      \begin{subfigure}[b]{0.235\textwidth}
        \centering
        \includegraphics[width=1\textwidth]{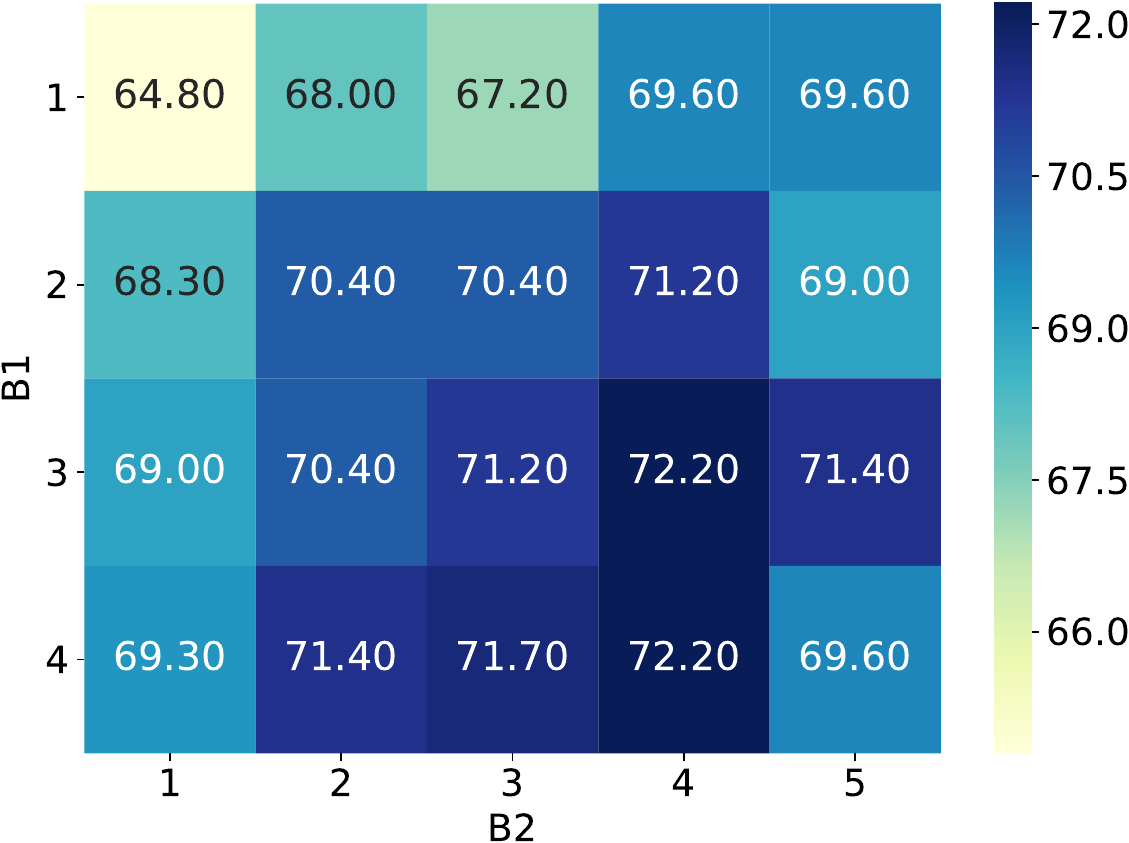}
        \caption{DSC on MBPP Plus.}
      \end{subfigure}
      \hfill
      \begin{subfigure}[b]{0.235\textwidth}
        \centering
        \includegraphics[width=1\textwidth]{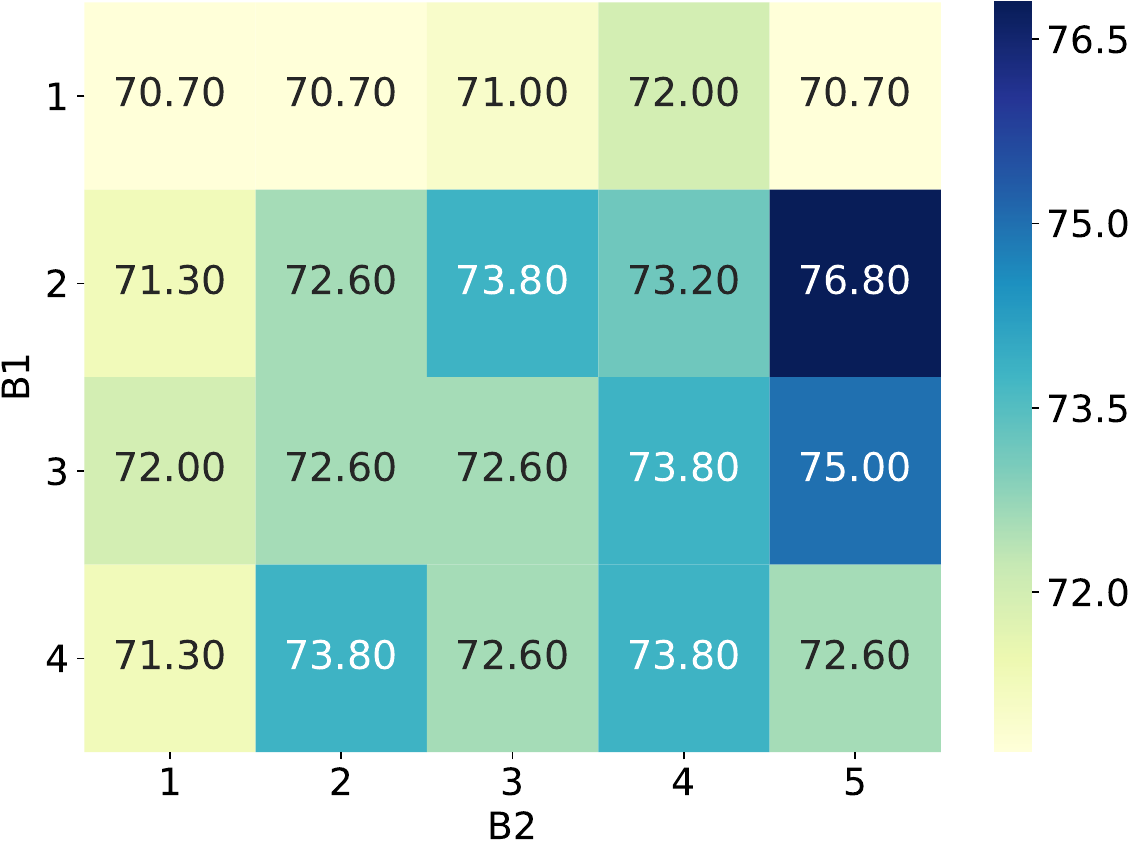}
        \caption{DSC on HumanEval Plus.}
      \end{subfigure}
      \hfill
      \begin{subfigure}[b]{0.235\textwidth}
        \centering
        \includegraphics[width=1\textwidth]{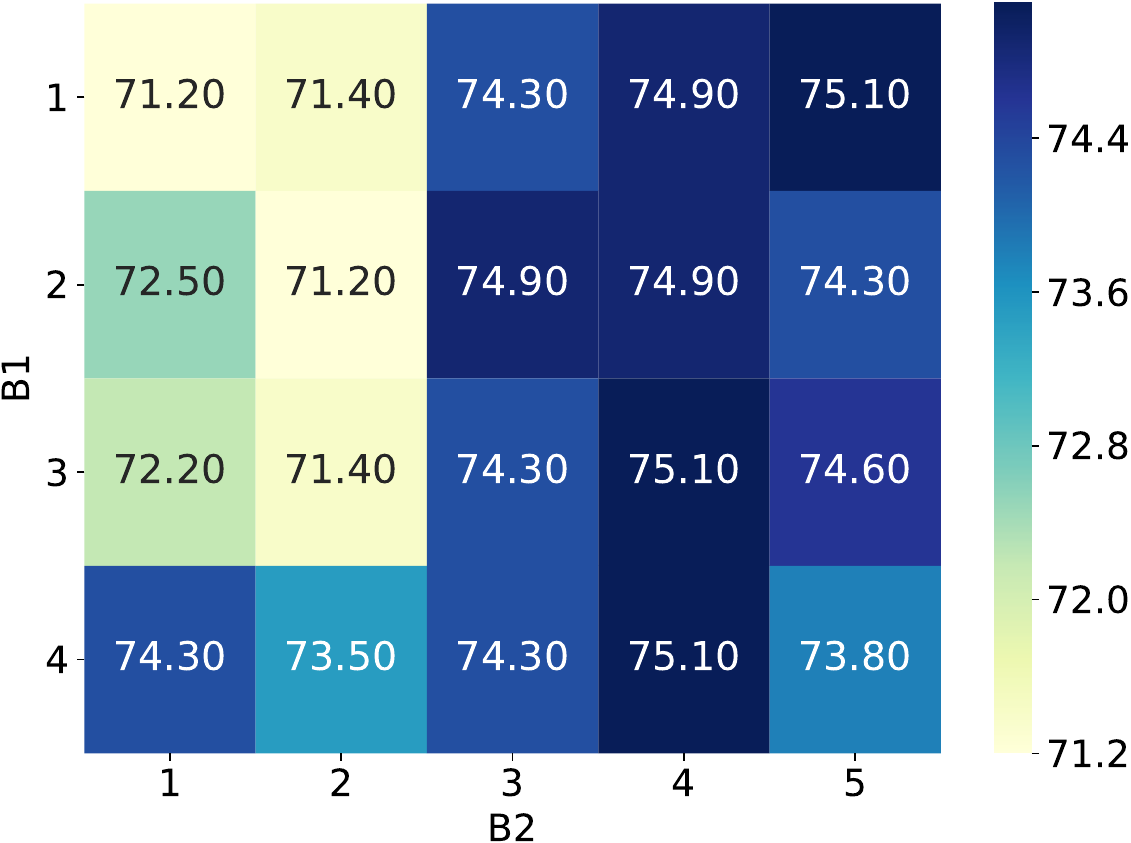}
        \caption{QC on MBPP Plus.}
      \end{subfigure}
      \hfill
      \begin{subfigure}[b]{0.235\textwidth}
        \centering
        \includegraphics[width=1\textwidth]{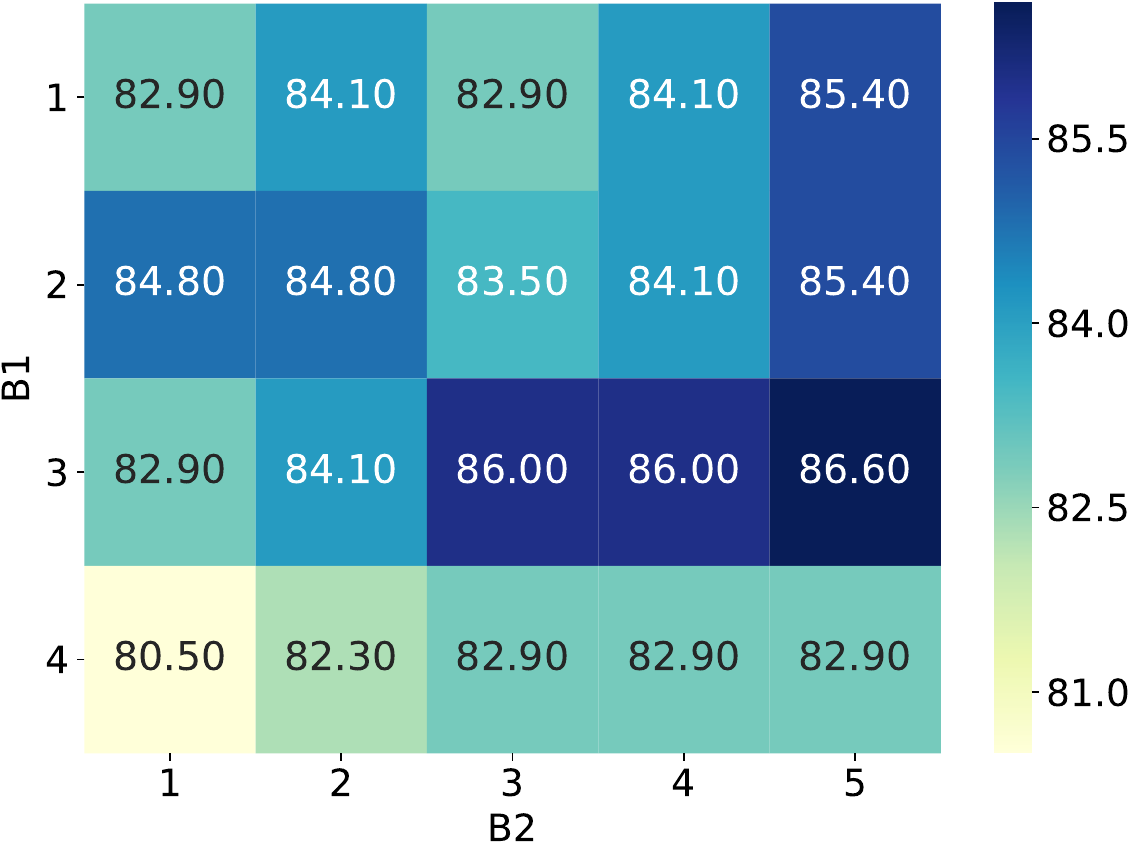}
        \caption{QC on HumanEval Plus.}
      \end{subfigure}
    \caption{Parameter analysis of $B_1$ and $B_2$.
   } 
    \vspace{-0.3cm}
	\label{fig:parameter1}
\end{figure*}

\textbf{Adaptive CoT reasoning.} Finally, we validate the effectiveness of the adaptive CoT reasoning mechanism by disabling direct solution generation in line 7 in Algorithm~\ref{algo:sbs}. The performance decreases are substantial, with DeepSeek-Coder showing a \gsz{3.6\%}  
drop on HumanEval Original and a \gsz{3.2\%}  
drop on MBPP Plus. For Qwen2.5-Coder, we observe a \gsz{3.2\%} 
reduction on HumanEval Original and a \gsz{1.7\%} 
decrease on MBPP Plus. Furthermore, removing adaptive CoT reasoning also increases average code generation time. For example, on MBPP, \tool enables to only apply step-by-step reasoning for 65.6\% and 73.5\% of questions for \gsz{DeepSeek-Coder} and \gsz{Qwen2.5-Coder}, respectively. Removing this component increases generation time for these two models by 34.5\% and 32.6\%.
These results indicate that adaptive CoT reasoning effectively contributes to choosing the appropriate generation approach and reducing unnecessary computational cost.


\begin{figure}[t]
     \centering
     \begin{subfigure}[h]{0.23\textwidth}
        \centering
        \includegraphics[width=1 \textwidth]{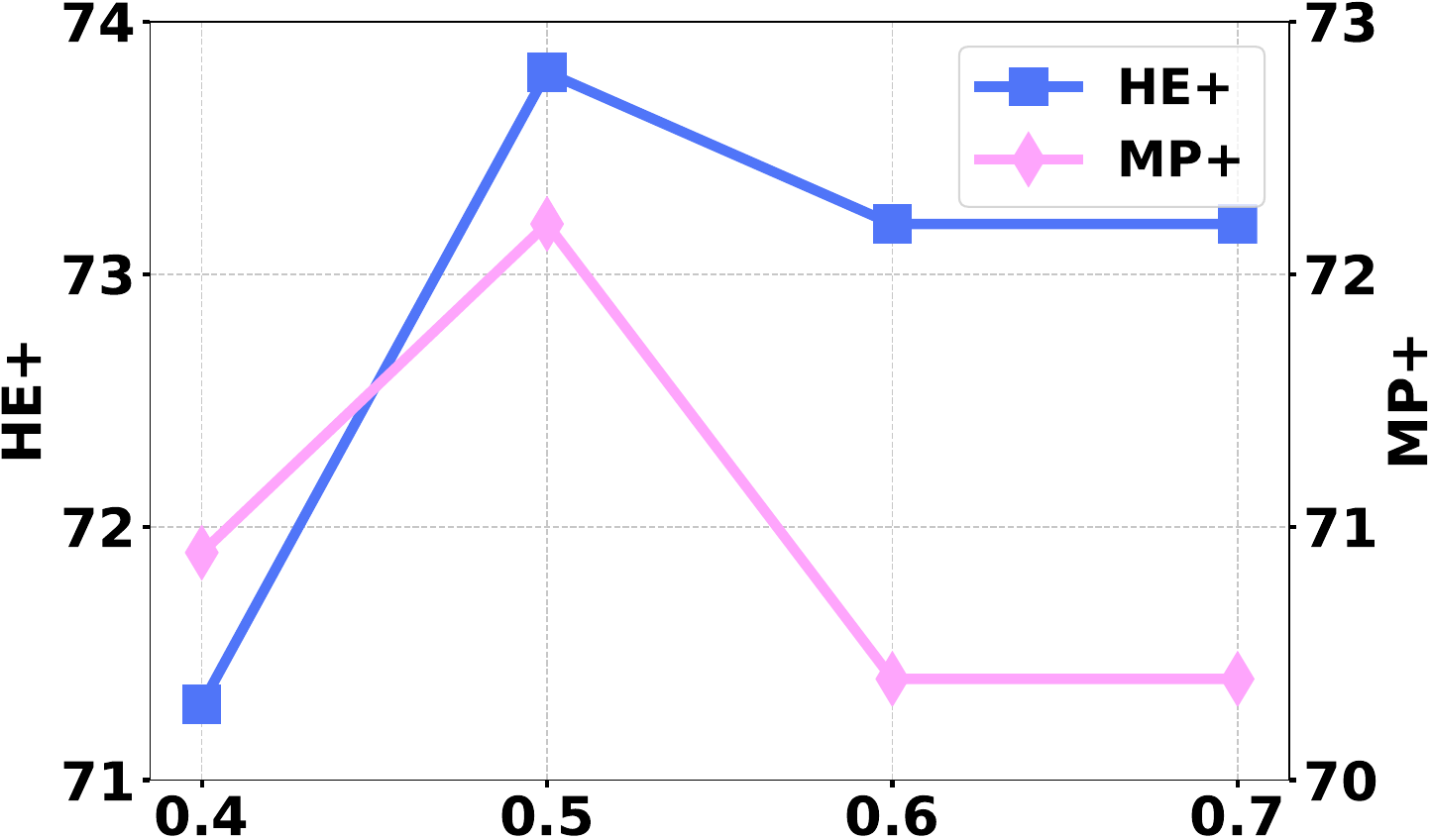}
        \caption{Temperature on DSC.}
    \end{subfigure}
    \hfill
    \begin{subfigure}[h]{0.23\textwidth}
        \centering
        \includegraphics[width=1 \textwidth]{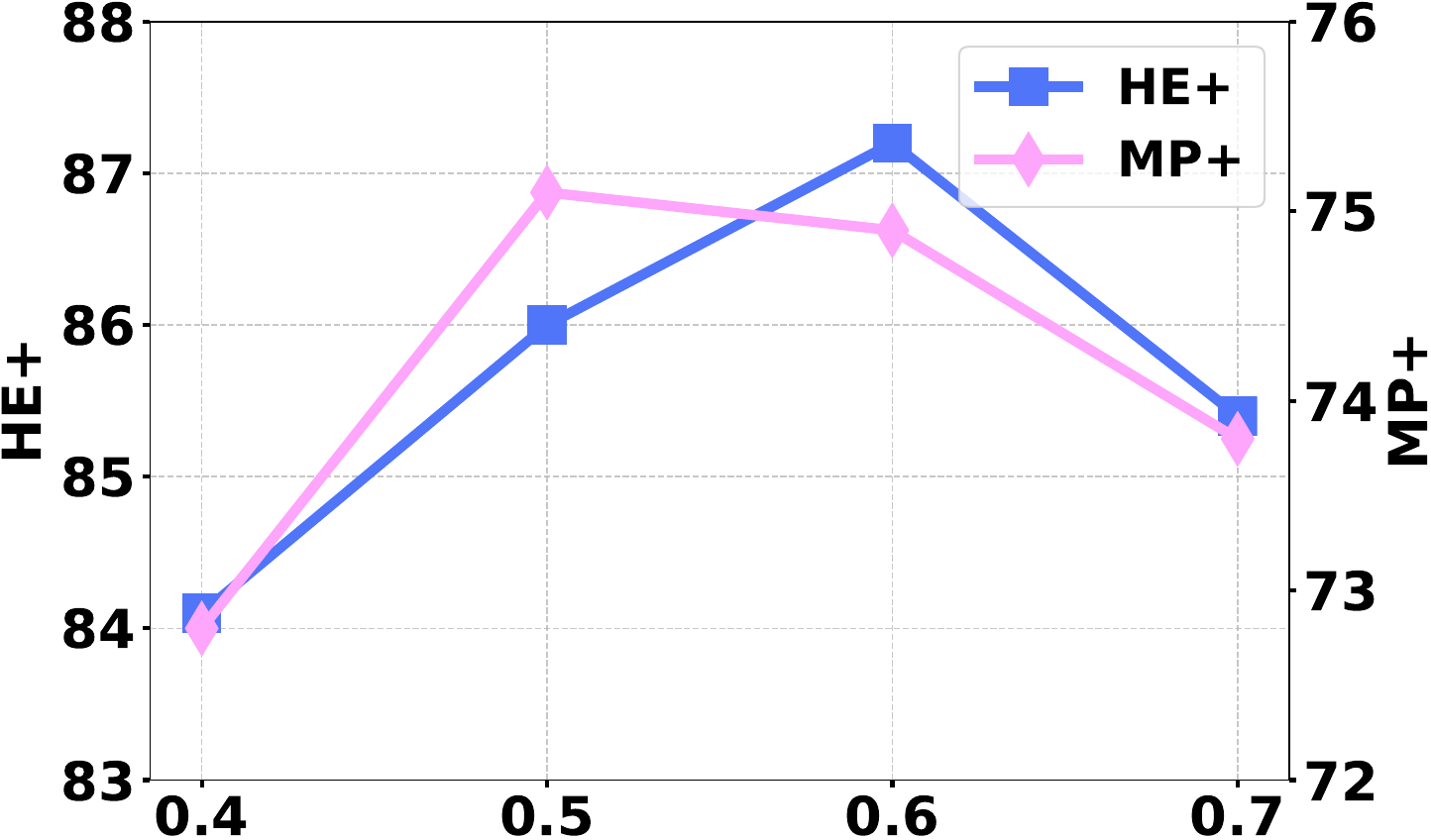}
        \caption{Temperature on QC.}
    \end{subfigure}
    \begin{subfigure}[h]{0.23\textwidth}
        \centering
        \includegraphics[width=1 \textwidth]{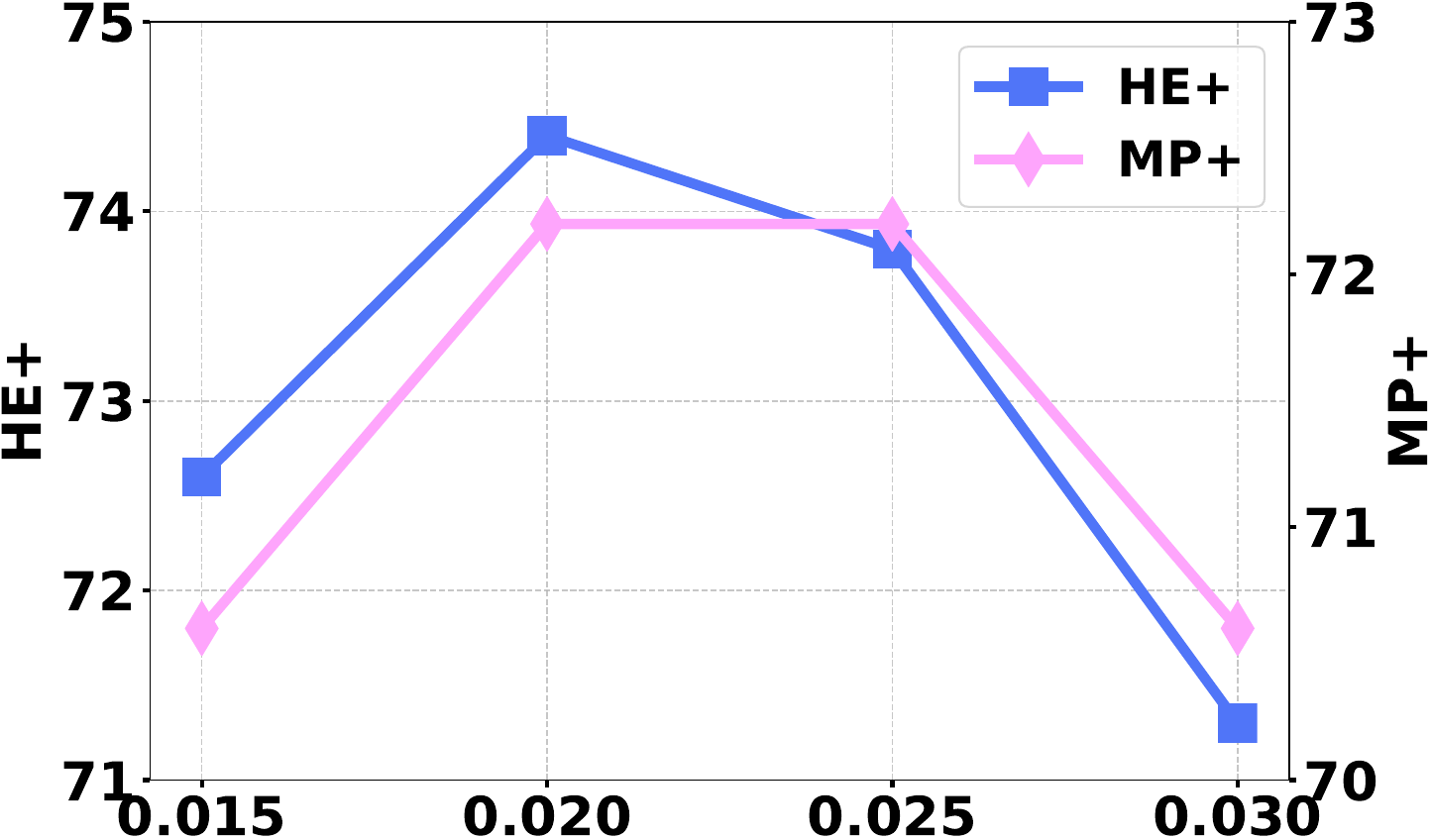}
        \caption{$\beta$ on DSC.}
    \end{subfigure}
    \hfill
    \begin{subfigure}[h]{0.23\textwidth}
        \centering
        \includegraphics[width=1 \textwidth]{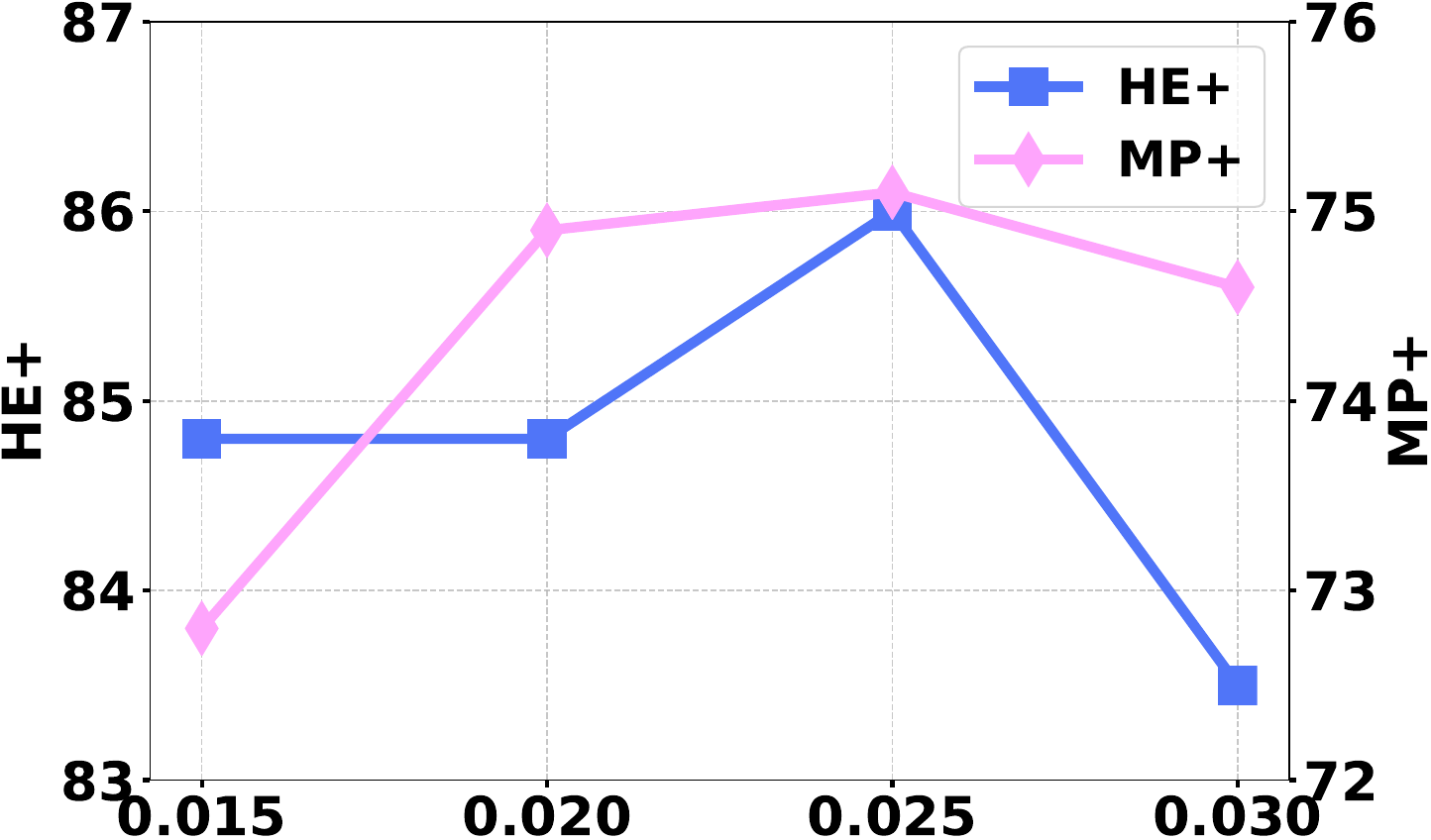}
        \caption{$\beta$ on QC.}
    \end{subfigure}        
    \caption{Parameter analysis of temperature and value model weight $\beta$.
   }  
     \label{fig:parameter}
    \vspace{-0.5cm}
\end{figure}

\subsubsection{Impact of different training strategy.}
In this section, we validate the effectiveness of our model training strategy. \tool consists of two training stages: (1) an initial phase using data collected during diverse reasoning path exploration, followed by (2) adaptive CoT training with modified objectives. 
To validate this design, 
we construct
two variants of our training strategy, including using only the first phase without the adaptive CoT training and maintaining the two-phase approach but without the KL divergence modification in the second phase.

\textbf{Only First Stage.} When limiting \tool to only the first stage of training, we observe performance degradation across all benchmarks. For example, with \gsz{DeepSeek-Coder,} this variant achieves \gsz{80.4\%} 
on HumanEval Original and \gsz{71.8\%} 
on HumanEval Plus, representing a notable decline compared to the 
\tool. This suggests that the absence of adaptive CoT reasoning training impairs the model's ability to assess reasoning quality, thus hampering its capacity to effectively select between step-by-step reasoning and direct generation approaches.


\textbf{Two Stage no KL.} The two-stage approach without object modification shows remarkable performance degradation across all tested models. For example, on LiveCodeBench, this variant
performs \gsz{4.3\%} 
worse than \tool with \gsz{DeepSeek-Coder} and 13.1\% 
worse with Qwen. These results demonstrate that without KL regularization, models suffer from severe catastrophic forgetting, degrading their original generation capabilities.

 \begin{tcolorbox}[breakable,width=\linewidth,boxrule=0pt,top=1pt, bottom=1pt, left=1pt,right=1pt, colback=gray!20,colframe=gray!20]
 \textbf{Answer to RQ2:}
 All components contribute to \tool for code generation. 
 Our two-stage training strategy demonstrates the best performance among different variants.
 \end{tcolorbox}

\subsection{RQ3: Parameter Analysis}\label{sec:para}

In this section, we study the impact of four parameters on the performance of \tool, including the LLM temperature, the weight of value model $\beta$ in Equ.~\ref{eq:1}, and the beam search width $B_1$ and $B_2$ in the Algorithm~\ref{algo:sbs}. Due to space limitation, we only present the results on HumanEval Plus and MBPP Plus, with results for other benchmarks presented on our GitHub repository~\cite{replicate}. 

\textbf{Temperature.} To investigate the impact of temperature on model performance, we vary the temperature from 0.4 to 0.7 and display the results in Figures~\ref{fig:parameter} (a) and (b). For \gsz{DeepSeek-Coder,} performance on both HumanEval Plus and MBPP Plus datasets achieves optimal results when the temperature is set to 0.5. In the Qwen2.5-Coder evaluation, the performance on HumanEval Plus reaches its peak at a temperature of 0.6, while MBPP Plus performs best at 0.5. In general, either lower (0.4) or higher (0.7) temperature values cannot yield better results. Based on these findings, we set the temperature parameter to 0.5 to obtain relatively stable and excellent performance across different models and datasets.


\textbf{Value model weight $\beta$.} We study the effect of $\beta$ by adjusting its value from 0.015 to 0.030, as shown in Figure~\ref{fig:parameter} (c) and (d). For \gsz{DeepSeek-Coder,} performance on HumanEval Plus peaks at a weight of 0.020, while performance on MBPP Plus performs best at 0.025. For \gsz{Qwen2.5-Coder}, performance on both HumanEval Plus and MBPP Plus datasets achieves optimal results at 0.025. When the weight is too small, the value model may not learn effectively to assess the quality of the reasoning paths. In contrast, when the weight increases to 0.030, performance decreases on both datasets across both evaluation models, suggesting that excessively high weight values may impair the generation capabilities of the policy model. Therefore, we chose to set the weight parameter to 0.025 to achieve balanced performance across different encoder models and evaluation datasets.

\textbf{Beam size $B_1$ and $B_2$.} Figure~\ref{fig:parameter1} presents the influence of $B_1$ and $B_2$ on model performance. We systematically increase the values of both parameters from (1,1) and monitor the resulting performance. For \gsz{DeepSeek-Coder,} we observe a gradual improvement in performance on HumanEval Plus as shown in Figure~\ref{fig:parameter1} (a), with performance increasing until approximately $B_1$=3 and $B_2$=4. A similar trend is evident in Figure~\ref{fig:parameter1} (b) for DSC on MBPP Plus, where the optimal values cluster around the similar region. For \gsz{Qwen2.5-Coder}, Figures~\ref{fig:parameter1} (c) and (d) demonstrate comparable behavior, with performance metrics stabilizing around $B_1$=3 and $B_2$=4 for both HumanEval Plus and MBPP Plus datasets. Although further increasing these parameters might yield marginal improvements, the additional computational cost would not justify such minimal gains. Therefore, we select $B_1$=3 and $B_2$=4 as our final parameter values.

 \begin{tcolorbox}[breakable,width=\linewidth,boxrule=0pt,top=1pt, bottom=1pt, left=1pt,right=1pt, colback=gray!20,colframe=gray!20]
 \textbf{Answer to RQ3:} 
The LLM temperature and value model weight $\beta$ could affect the accuracy of \tool, with intermediate values yielding optimal performance. Increasing beam sizes $B_1$ and $B_2$ brings substantial improvement, though gains become relatively marginal beyond $B_1$=3 and $B_2$=4.
 \end{tcolorbox}

\section{Discussion}\label{sec:discuss}


\subsection{Why \tool works for code generation?}\label{sec:case}
\begin{figure}[t]
    \centering
    \includegraphics[width=1\linewidth]{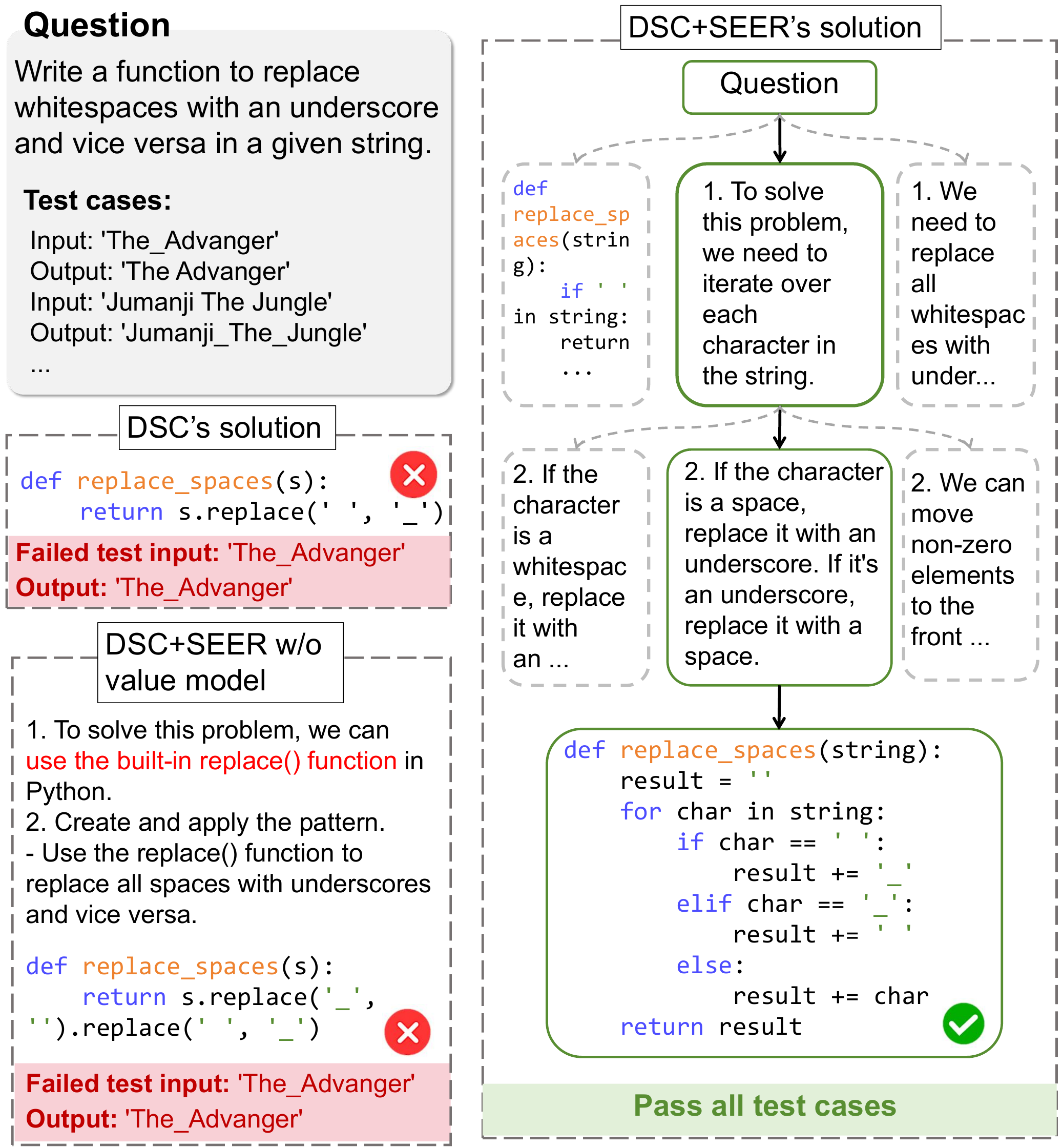}
    \caption{Case studies of comparing \tool with baseline methods. 
    The dashed lines in DSC+\tool's solution represent unselected steps.}
    \label{fig:case}
    \vspace{-0.5cm}
\end{figure}

To better understand how \tool helps CoT reasoning for code generation, we present an example from the MBPP benchmark in Figure~\ref{fig:case}. As demonstrated, the baseline DeepSeek-Coder directly generates a solution by using a single \texttt{replace()} function, which fails to handle the bidirectional replacement requirement in the test case ``The\_Advanger''. This illustrates a fundamental limitation of direct generation approaches, where models often lack comprehensive requirement analysis and overlook complex edge cases. In contrast, our \tool approach allows DeepSeek-Coder to break down the problem through step-by-step reasoning, leading to a character-by-character iteration solution that correctly handles both space-to-underscore and underscore-to-space conversions. However, when the value model-based reasoning quality assessment is removed and the traditional beam search is used instead, the model produces flawed reasoning from the very first step by attempting to use the \texttt{replace()} function. Although this solution is closer to the correct answer than the baseline, its reasoning process overlooks the coverage problem of two consecutive transformations. This indicates that value model-based reasoning quality assessment plays a crucial role in guiding the reasoning process.

\subsection{\gsz{Relationship with Large Reasoning Models}}

\gsz{Recently, Large Reasoning Models like OpenAI-O3~\cite{o3} and DeepSeek-R1~\cite{guo2025deepseek} have gained attention for using ``long CoT'' to achieve test-time scaling~\cite{DBLP:journals/corr/abs-2503-24235,DBLP:journals/corr/abs-2504-13828,wen2025vul}. Unlike traditional CoT's linear, step-by-step generation process, long CoT allows models to iteratively refine their thoughts through activities like reflection and backtracking~\cite{guo2025deepseek}. \tool is a form of test-time scaling based on tree search, which is orthogonal to the long CoT-based method. Tree search methods generate multiple candidate steps and use value models to select the most promising ones, ensuring step-wise reliability~\cite{DBLP:journals/corr/abs-2504-13828}. In contrast, long CoT employs linear reasoning to continuously revise content, offering broader domain applicability without requiring auxiliary scoring models. Consequently, long CoT can suffer from ``overthinking'' and inefficiency issue due to extended generation times~\cite{DBLP:journals/corr/abs-2504-13828,DBLP:journals/corr/abs-2502-03373}. Meanwhile, recent research~\cite{DBLP:journals/corr/abs-2504-07986,DBLP:journals/corr/abs-2504-12329} also indicates that these two approaches can be further combined to achieve better performance. For example, \cite{DBLP:journals/corr/abs-2504-07986} has shown that LRMs naturally use certain symbols like ``Wait'' to distinguish different steps, so these multiple steps can be viewed as a decision-making process where MCTS can be used to construct step-by-step process supervision data. Our proposed method can then be employed to enrich the diversity of training data and assist efficient model reasoning.}

\gsz{We further conduct experiments to compare \tool with a representative open-source LRMs R1-Distill-Qwen-7B~\cite{guo2025deepseek}, with respect to both accuracy and generation time. Comparing DeepSeek-R1-Distill-Qwen-7B with \tool is not entirely fair, as it requires data distilled from a proprietary model, whereas \tool does not. However, the experimental results demonstrate that our method achieves comparable overall performance. Specifically, our approach outperforms DeepSeek-R1-Distill-Qwen-7B on MBPP with a pass rate by 7.4\%. More importantly, our method consistently requires significantly less inference time across all benchmarks. Based on the average results across various datasets, our method achieves 75.6\% compared to 73.0\% for DeepSeek-R1-Distill-Qwen-7B, while maintaining time efficiency advantages. This demonstrates the effectiveness of \tool.}

\begin{table*}[t]
    \centering
    \caption{\gsz{Comparison with DeepSeek-R1-Distill-Qwen-7B.}}
    \vspace{-0.3cm}
    \scalebox{0.85}{
    \begin{tabular}{l|ccc|ccc|cc|c}
    \toprule
     \multirow{2}{*}{Methods}  & \multicolumn{3}{c|}{MBPP} & \multicolumn{3}{c|}{HumanEval} & \multicolumn{2}{c|}{LiveCodeBench} & \multirow{2}{*}{Average}\\
         & Pass rate & Plus & time & Pass rate & Plus & time & Pass rate & time & \\
    \midrule
    Qwen2.5-Coder-7B-Instruct+SEER & \textbf{88.6} & \textbf{75.1} & 7min40s & 90.9 & \textbf{86.0} & 5min20s & 36.7 & 73min & 75.6 \\
    DeepSeek-R1-Distill-Qwen-7B & 81.2 & 67.7 & 12min49s & \textbf{91.5} & 84.1 & 6min09s & \textbf{40.5} & 99min & 73.0 \\
    \bottomrule
    \end{tabular}
    }
    \vspace{-0.3cm}
    \label{tab:r1_comparison}
\end{table*}

\subsection{\gsz{Relationship with Reflexion}}
\gsz{In this section, we further validate whether \tool could be combined with existing reasoning frameworks like Reflexion~\cite{DBLP:conf/nips/ShinnCGNY23}. Reflexion is a framework that iteratively refines code generation through self-reflection and error correction. \tool aims to enhance code generation accuracy in the first attempt while Reflexion uses iterative generation and validation which is orthogonal to our method. The experimental results in Table~\ref{tab:reflextion} demonstrate that combining \tool with Reflexion consistently improves performance across all evaluated benchmarks and model sizes. For the Deepseek-Coder, \tool+Reflexion achieves notable improvements on MBPP Plus with 3.5 percentage points gain from 72.2\% to 75.7\% and on LCB average with 1.9 percentage points improvement from 23.8\% to 25.7\%. Similarly, for the Qwen2.5-Coder, \tool+Reflexion demonstrates consistent gains across all benchmarks, with improvements ranging from 0.6 to 1.6 percentage points. These results indicate the complementary benefits of our approach with existing self-reflection techniques for code generation.}

\begin{table}[t]
    \centering
    \caption{\gsz{Results of combining \tool and Reflexion.}}
    \vspace{-0.3cm}
    \scalebox{0.85}{
    \begin{tabular}{l|cc|cc|c}
    \midrule
    \multirow{2}{*}{Approach}  & \multicolumn{2}{c|}{MBPP} & \multicolumn{2}{c|}{HumanEval} & LCB \\
    & Original & Plus & Original & Plus & Avg \\
    \midrule
    \multicolumn{6}{c}{Deepseek-Coder-6.7B} \\
    \midrule
    \tool & 82.0 & 72.2 & 82.9 & 73.8 & 23.8 \\
    \tool+Reflexion & \textbf{83.9} & \textbf{75.7} & \textbf{82.9} & \textbf{74.4} & \textbf{25.7} \\
    \midrule
    \multicolumn{6}{c}{Qwen2.5-Coder-7B-Instruct} \\
    \midrule
    \tool & 88.6 & 75.1 & 90.9 & 86.0 & 36.7 \\
    \tool+Reflexion & \textbf{88.9} & \textbf{77.0} & \textbf{91.5} & \textbf{86.6} & \textbf{38.3} \\
    \bottomrule
    \end{tabular}
    }
    \vspace{-0.3cm}
    \label{tab:reflextion}
\end{table}

\subsection{\gsz{Performance on Other Programming Languages}}
\gsz{To assess \tool's generalizability across different programming languages, we evaluate it on C++ and Java from the MBPP portion of the MultiPL-E~\cite{cassano2022multipl} benchmark using models trained on Python data. The results in Table~\ref{tab:cross_language} demonstrate that \tool consistently improves performance across both target languages, with notable gains on Java tasks. \tool achieves improvements of 0.76 and 4.41 points for Deepseek-Coder on C++ and Java respectively, and 1.01 and 2.07 points for Qwen2.5-Coder, respectively. These results confirm that \tool's enhanced reasoning capabilities can successfully transfer to other programming languages even when trained only on Python data. In the future, we plan to extend \tool's training to include additional programming languages to further validate its versatility.}

\subsection{Threats to Validity}
We have identified the following major threats to validity:
\begin{enumerate}
\item 
\textit{Limited LLMs.} In this work, we select two state-of-the-art pre-trained models Deepseek-Coder and Qwen2.5-Coder for evaluation. To comprehensively evaluate the performance of \tool, we will validate whether \tool is also effective on other models such as Llama~\cite{DBLP:journals/corr/abs-2407-21783} in the future. 


\item 
\textit{Potential data contamination.} Another threat is potential data contamination. To address this, \gsz{we utilize the Self-OSS dataset as our seed data, where the original authors have already performed comprehensive data decontamination and removed benchmark questions with high similarity.} \gsz{Additionally,} we conduct experiments on LiveCodeBench, a contamination-free benchmark that collects new problems over time. 


\end{enumerate}



\section{Related work}\label{sec:related}

\subsection{Code Generation}
Code generation aims to automatically create software code based on user intent. Recent advances in deep learning techniques, particularly LLMs, have significantly advanced this field. Several code LLMs trained on massive code datasets have achieved promising results, including StarCoder~\citep{DBLP:journals/corr/abs-2402-19173}, Code Llama~\citep{DBLP:journals/corr/abs-2308-12950}, DeepSeek-Coder~\citep{guo2024deepseek} and Qwen2.5-coder~\citep{hui2024qwen2}. Apart from training the code LLM, recent work has also focused on improving code generation performance through iterative generation and fixing approaches. For example, Dong et al.~\citep{DBLP:journals/tosem/DongJJL24} propose a self-collaboration method that enhances LLMs' performance by employing multiple LLMs as different roles in software development. Self-edit~\cite{DBLP:conf/acl/ZhangLLLJ23} utilizes compiler error messages to enhance the correctness of code generation. Our approach focuses on training the code LLMs to improve its reasoning ability and enhance code generation accuracy on the first attempt, which could complement these methods to further improve their performance.

\begin{table}[t]
    \centering
    \caption{\gsz{Performance on C++ and Java dataset.}}
    \vspace{-0.3cm}
    \scalebox{0.85}{
    \begin{tabular}{l|cc}
    \toprule
     Methods  & C++ & Java \\
    \midrule
    Deepseek-Coder-6.7B-Instruct & 63.22 & 63.21 \\
    Deepseek-Coder-6.7B-Instruct+\tool & 63.98 & 67.62 \\
    Qwen2.5-Coder-7B-Instruct & 63.73 & 72.80 \\
    Qwen2.5-Coder-7B-Instruct+\tool & 64.74 & 74.87 \\
    \bottomrule
    \end{tabular}
    }
    \vspace{-0.3cm}
    \label{tab:cross_language}
\end{table}
\subsection{LLM Reasoning}
Large language models (LLMs) have demonstrated the emergent ability to generate intermediate reasoning steps for complex problem-solving which is known as chain-of-thought reasoning~\citep{DBLP:conf/nips/Wei0SBIXCLZ22,DBLP:conf/nips/KojimaGRMI22}. To leverage this capability, researchers have proposed various prompting methods including Least-to-Most prompting~\citep{DBLP:conf/iclr/ZhouSHWS0SCBLC23}, Program-of-Thoughts (PoT)~\citep{DBLP:journals/tmlr/ChenM0C23}, Self-Planning~\citep{DBLP:journals/tosem/JiangDWFSLJJ24}, and Self-Refine~\citep{DBLP:conf/nips/MadaanTGHGW0DPY23}. Recent studies have also explored enhancing LLM reasoning through supervised fine-tuning~\citep{DBLP:conf/nips/ZelikmanWMG22,DBLP:journals/corr/abs-2406-03816}. Due to the high cost of manual data annotation, researchers have developed methods that synthesize reasoning process data by LLMs. For instance, the Self-Taught Reasoner (STaR)~\citep{DBLP:conf/nips/ZelikmanWMG22} generates rationales and fine-tunes on those that lead to correct answers. For complex multi-step reasoning tasks like mathematical and commonsense reasoning, some approaches employ MCTS to collect process supervision data for model training~\citep{DBLP:journals/corr/abs-2405-03553,DBLP:journals/corr/abs-2405-00451,DBLP:journals/corr/abs-2406-03816}. Different from those work, our work present as the first work to improve LLM's reasoning ability for code generation and improves the MCTS-based data collection through path perturbation and refinement to increase data diversity.


\section{Conclusion}\label{sec:conclusion}
In this paper, we introduce \tool, a novel framework that formulates code generation as a decision-making problem with diverse reasoning path exploration, reasoning quality-aware model training, and adaptive CoT reasoning. \tool effectively addresses the key limitations of previous work: limited exploration of reasoning paths, lack of quality assessment for the intermediate reasoning step, and the negative impact of ``overthinking''. Experimental results demonstrate that \tool effectively improves code generation performance across multiple benchmarks, enabling smaller models to outperform larger ones.




\bibliographystyle{ACM-Reference-Format}
\bibliography{sample}
\end{document}